\documentclass[manuscript]{acmart}

\usepackage{graphicx}
\usepackage{caption,subcaption}
\usepackage{microtype}
\usepackage{color,soul}
\usepackage{multirow}
\usepackage{tabularx}
\usepackage{booktabs}

\AtBeginDocument{%
  \providecommand\BibTeX{{%
    \normalfont B\kern-0.5em{\scshape i\kern-0.25em b}\kern-0.8em\TeX}}}

\setcopyright{acmcopyright}
\copyrightyear{2023}
\acmYear{2023}

\begin{document}

\title[Justification vs. Transparency]{Justification vs. Transparency: \textit{Why} and \textit{How} Visual Explanations in a Scientific Literature Recommender System}


\author{Mouadh Guesmi}
\affiliation{%
  \institution{University of Duisburg-Essen}
  \streetaddress{Forsthausweg 2}
  \city{Duisburg}
  \country{Germany}}
\email{mouadh.guesmi@stud.uni-due.de}

\author{Mohamed Amine Chatti}
\affiliation{%
  \institution{University of Duisburg-Essen}
  \city{Duisburg}
  \country{Germany}}
\email{mohamed.chatti@uni-due.de}

\author{Shoeb Joarder}
\affiliation{%
  \institution{University of Duisburg-Essen}
  \city{Duisburg}
  \country{Germany}}
\email{shoeb.joarder@uni-due.de}

\author{Qurat Ul Ain}
\affiliation{%
  \institution{University of Duisburg-Essen}
  \city{Duisburg}
  \country{Germany}}
\email{qurat.ain@stud.uni-due.de}

\author{Clara Siepmann}
\affiliation{%
 \institution{University of Duisburg-Essen}
 \city{Duisburg}
 \country{Germany}}
\email{clara.siepmann@uni-due.de}

\author{Hoda Ghanbarzadeh}
\affiliation{%
 \institution{University of Duisburg-Essen}
 \city{Duisburg}
 \country{Germany}}
\email{hoda.ghanbarzadeh@stud.uni-due.de}

\author{Rawaa Alatrash}
\affiliation{%
  \institution{University of Duisburg-Essen}
  \city{Duisburg}
  \country{Germany}}
\email{rawaa.alatrash@stud.uni-due.de}

\renewcommand{\shortauthors}{Guesmi et al.}

\begin{abstract}
Significant attention has been paid to enhancing recommender systems (RS) with explanation facilities to help users make informed decisions and increase trust in and satisfaction with the RS. Justification and transparency represent two crucial goals in explainable recommendation. Different from transparency, which faithfully exposes the reasoning behind the recommendation mechanism, justification conveys a conceptual model that may differ from that of the underlying algorithm. An explanation is an answer to a question. In explainable recommendation, a user would want to ask questions (referred to as intelligibility types) to understand results given by the RS. In this paper, we identify relationships between \textit{Why} and \textit{How} explanation intelligibility types and the explanation goals of \textit{justification} and \textit{transparency}. We followed the Human-Centered Design (HCD) approach and leveraged the What-Why-How visualization framework to systematically design and implement \textit{Why} and \textit{How} visual explanations in the transparent Recommendation and Interest Modeling Application (RIMA). Furthermore, we conducted a qualitative user study (N=12) to investigate the potential effects of providing \textit{Why} and \textit{How} explanations together in an explainable RS on the users' perceptions regarding transparency, trust, and satisfaction. Our study showed qualitative evidence confirming that the choice of the explanation intelligibility types depends on the explanation goal and user type.   
\end{abstract}


\keywords{Recommender Systems, Explainable Recommendation, Visualization, Transparency, Justification, Trust}

\maketitle
\section{Introduction}
Recommender systems (RS) have become an integral part of our daily lives, assisting users in discovering relevant items or services in various domains, such as e-commerce, social media, and entertainment. The success of a RS depends on its ability to accurately predict user preferences and recommend relevant items. However, as the underlying algorithms become more complex, the transparency and interpretability of the system decrease. Current RS often appear as “black boxes” by hiding important details from their users. As a consequence, users may not understand the system’s behavior and create an unfitting mental model of the RS, especially if the system behaves unexpectedly, causing a lack of confidence among users who may then lose trust, get frustrated, and eventually reject the system’s recommendations \cite{tintarev2015explaining,nunes2017systematic,zhang2020explainable,kunkel2019let}. Hence, research increasingly has taken user-centric aspects such as transparency, trust, and user satisfaction with an RS into account, when assessing its quality \cite{pu2012evaluating,knijnenburg2012explaining,konstan2012recommender}. 
The lack of transparency in many AI systems and specifically recommendation techniques has sparked interest in incorporating explanation in RS, with the goal of making these RS more transparent and providing users with information that can aid in the development of an accurate mental model of the system's behavior. This aspect is also important for RS providers to build and maintain user trust in the system. Therefore, equipping RS with explanation benefits both users and system designers \cite{tintarev2015explaining,nunes2017systematic,zhang2020explainable}. 

Explainable recommendation refers to personalized recommendation algorithms that not only provide users or system designers with recommendation results, but also explanations to clarify the reason for such items to be recommended. Explanations are a necessary condition to help users build an accurate mental model of the RS. Generally, an explanation seeks to answer questions, also called intelligibility types, such as \textit{What}, \textit{Why}, \textit{How}, \textit{What if}, and \textit{Why not} \cite{lim2009assessing} in order to achieve understanding. \citet{lim2013evaluating} found that users may exploit different strategies to understand AI systems and thus use different intelligibility types for the different explanation goals. 

 Primary goals of explainable recommendation include transparency, effectiveness, trust, persuasiveness, efficiency, scrutability, and debugging \cite{tintarev2007survey,jannach2019explanations}. Transparency is a crucial goal that explanations can serve. It refers to exposing (parts of) the reasoning behind the recommendation mechanism to explain how the system works \cite{tintarev2015explaining}. \textit{Transparency} is closely related to \textit{justification}. There is, however, an important distinction between the two concepts. While transparency focuses on explaining the RS \textit{process} and provides detailed insights into \textit{how} the RS works, justification focuses on the RS \textit{output} and merely gives a plausible abstract description that might be decoupled from the recommendation algorithm to answer the question of \textit{why} items have been recommended without revealing the inner working of the RS \cite{tintarev2012evaluating,tintarev2015explaining,balog2019,vig2009tagsplanations}. 

In this work, we focus on the justification and transparency goals in explainable RS by providing \textit{Why} and \textit{How} explanations in an explainable RS. We are particularly interested in how to design and implement visual explanations. Visualizations are a popular medium to provide insight into data or how a system works \cite{lim2009why}. In general, humans can process visual information faster and much easier as compared to textual information \cite{munzner2014visualization}. Recognizing their benefits, visualizations are increasingly used to deliver explanations in RS \cite{herlocker2000explaining,gedikli2014,nunes2017systematic,zhang2020explainable,guesmi2021demand}.

It has been shown in some works that users may benefit from \textit{How} explanations while other works uncover that there are circumstances when these detailed explanations are not always beneficial and that \textit{Why} explanations are often enough to help users understand the recommendations \cite{herlocker2000explaining,vig2009tagsplanations,gedikli2014}. It is thus
important to provide explanations with enough details to allow users to build accurate mental models of how the RS operates without overwhelming them. However, in the existing literature on explainable recommendation, significant gaps remain when it comes to understanding when and if \textit{Why} and \textit{How} explanations are necessary or useful. Although there are studies addressing \textit{Why} and \textit{How} explanations in RS, to the best of our knowledge, there is no prior research that integrates both explanations side-by-side in the same RS. Moreover, little attention has been paid to how to systematically design \textit{Why} and \textit{How} explanations in RS, as well as how these explanations would affect user perception of transparency, trust, and satisfaction with the RS when they are provided together. 

To address these research gaps, in this paper, we follow the Human-Centered Design (HCD) approach \cite{norman2013design} and leverage Munzner's What-Why-How visualization framework \cite{munzner2014visualization} to systematically design \textit{Why} and \textit{How} visual explanations and provide them together in the transparent Recommendation and Interest Modeling Application (RIMA) that gives explanations of recommended scientific publications. 
Further, we conducted a qualitative user study (N=12) based on moderated think-aloud sessions and semi-structured interviews with students and researchers to explore how users
perceive \textit{Why} and \textit{How} visual explanations in an explainable RS.  

The objective of the study was to answer the following research question \textbf{(RQ)}: What is the potential impact of \textit{Why} and \textit{How} visual explanations on
the user perceptions regarding transparency, trust, and user satisfaction, when these two explanations are provided together in an explainable RS? The results of our study provide qualitative evidence that: (1) It is important to differentiate between objective transparency and user-perceived transparency; (2) The user perceptions of \textit{Why} and \textit{How} explanations in terms of transparency and trust depend on the user type (e.g., background knowledge); (3) there is a trade-off "transparency/trust vs. satisfaction" when \textit{Why} and \textit{How} explanations are provided together in the RS; (4) The choice of the explanation intelligibility questions depends on the explanation goal.

In summary, this work makes the following three main contributions: 
First, we identify relationships between different intelligibility types (i.e., Why and How) and explanation goals (i.e., justification and transparency). 
Second, we systematically design \textit{Why} and \textit{How} explanations by following the HCD approach and the What-Why-How visualization framework and integrate both explanations in the same RS. 
Third, we investigate the potential impact of \textit{Why} and \textit{How} explanations on the perception of explainable recommendation in terms of transparency, trust, and user satisfaction.
\section{Related Work}
This section discusses related work on explainable recommendation in relation to the two explanation's goals of justification and transparency and the two intelligibility types \textit{Why} and \textit{How}. 
\subsection{Justification vs. Transparency}
System transparency is defined as the extent to which information about a system’s reasoning is provided and made available to users \cite{hosseini2018four,zhao2019users}. In the literature on AI and advice-giving systems (AGS), transparency is often linked to users’ understanding of systems’ inner logic \cite{pu2012evaluating}. \citet{zhao2019users} suggests two alternative ways of measuring system transparency, namely from systems’ perspective and users’ perspective. The authors further distinguish between objective transparency (i.e., the extent to which systems release information about how they work), subjective transparency (i.e., the extent to which users perceive such information is available), and users’ perceived transparency (i.e., the extent to which users feel that they understand the meaning of the provided information). Providing transparency is generally considered to be beneficial to users and could enhance users’ trust in the system, which in turn could increase
users’ acceptance of system's outcomes \cite{cramer2008effects,diakopoulos2017algorithmic,hosseini2018four,harman2014dynamics,kunkel2019let,ananny2018seeing,zhao2019users,pu2012evaluating}. However, various studies found that revealing too much detail about how the system's inner logic may result in information overload, confusion, and a low level of perceived understanding, which may in turn reduce users’ trust in and acceptance of the system \cite{tintarev2007survey,ananny2018seeing,hosseini2018four}. For lay-users, revealing the system's functionality at an abstract level would help them build an accurate mental model of the system, without overwhelming them. This suggests that there should be an optimal level of transparency which will generate the highest level of users’ perceived understanding of and trust in the system \cite{zhao2019users}. 

Transparency in RS is related to the capability of a system to expose the reasoning behind a recommendation to its users \cite{herlocker2000explaining}, and is defined as users’ understanding of the RS' inner logic \cite{tintarev2007survey,pu2012evaluating}. In the RS domain, \citet{gedikli2014} also differentiate between objective transparency and user-perceived transparency. Objective transparency means that the RS reveals the actual mechanisms of the underlying algorithm. On the other hand, user-perceived transparency is based on the users’ subjective opinion about how well the RS is capable of explaining its recommendations. User-perceived transparency can be high even though the RS does not actually reveal the underlying recommendation algorithm \cite{herlocker2000explaining}. In some cases (e.g., high complexity of the algorithm) and for some users, it might be more appropriate to justify a recommendation output instead of revealing the inner working of the RS \cite{gedikli2014}. Justification is the ability of the system to help the user understand why an item was recommended \cite{herlocker2000explaining}. These justifications are often more shallow and user-oriented \cite{vig2009tagsplanations}. Transparency and justification are closely tied together. However, transparency differs from justification in that while the former faithfully represents and exposes the reasoning about \textit{how} the recommendations are selected and \textit{how} the system works, the latter merely provides a plausible reason \textit{why} an item is recommended, that may be decoupled from the RS algorithm \cite{tintarev2007survey,tintarev2015explaining,vig2009tagsplanations,balog2019}. Often, the underlying algorithm is too complex or not intuitive to explain, or may involve details that the RS provider wishes to protect. Thus, current RS often opt for presenting more user-oriented justifications, rather than offering genuine transparency by explaining the rationale of the recommendation algorithm \cite{balog2019,gedikli2014,vig2009tagsplanations,zhang2020explainable}.

While many RS are considered as a “black box”, a transparent RS would also try to explain the reasoning behind a recommendation to the user \cite{gedikli2014}. Generally, an explanation in RS seeks to show how a recommended item relates to user’s preferences \cite{vig2009tagsplanations}. Two major goals that explanations can serve are to provide transparency (i.e., the ability to explain how the RS works \cite{tintarev2007survey}) or justification (i.e., the ability to justify the recommendations without revealing the actual mechanisms of the RS algorithm \cite{tintarev2012evaluating}). Justification is thus linked to post-hoc explanation, which aims at communicating understandable information about how an already developed model produces its predictions for any given input that may be decoupled from the initial model \cite{arrieta2020explainable}. For example, the recommendations of a black-box RS can be explained by making a post-hoc selection of the relevant features that have led to the recommendation, e.g., “We recommend you this because it has <features> you liked in the past” \cite{afchar2022explainability}.

At a high level, an explanation is an answer to a user's questions that leads to understanding \cite{miller2019explanation}. \citet{lim2009assessing} found that users ask a wide range of questions to understand AI systems. These questions, also called intelligibility queries or types, include Why, Why Not, How, What If, and How To \cite{lim2009assessing,liao2020questioning}. \citet{lim2013evaluating} found that users may exploit different strategies to understand AI systems and thus use different intelligibility types for the different explanation goals. 
\citet{lim2019these} described how to support three explanation goals (i.e., filter causes, generalize and learn, and predict and control) with the intelligibility explanation types and identified specific pathways mapping the use of these intelligibility types explicitly back to user goals. In the RS domain, pathways can also be established between intelligibility types and explanation goals. Specifically, \textit{How} explanations are often used to explain the rationale of the recommendation algorithm, and are thus related to \textit{transparency}. On the other hand, \textit{Why} explanations are used to present more abstract and user-oriented \textit{justification}.
\subsection{Why and How Explanations}
\textit{Why} explanations serve as a justification rather than a description of the underlying recommender algorithm. This intelligibility type is commonly used in the RS domain, as the basic goal that any explanation needs to support is to give a plausible reason why an item was recommended. In general, \textit{Why} explanations attempt to show why a recommended item fits one’s preferences, at an abstract level (e.g., "users who are similar to you also like..."). For instance, ‘Tagsplanation’ recommends movies and explains them based on similar movie content \cite{vig2009tagsplanations}. To justify the recommended movie based on community tags, ‘Tagsplanation’ uses bar charts to visualize each tag’s relevance in the movie and the user’s preference. 'HyPER' \cite{kouki2019personalized} is a music RS that explains recommended artists based on similarity with the artists in the user profile, artist’s popularity, and artists liked by similar users. 'HyPER' uses a Venn diagram to recommend intersecting music artists between sets of artists from a user profile, popular artists, and artists liked by people listening to similar artists as the user. 'Relevance Tuner+' \cite{tsai2019explaining} explains recommended conference talks and potential scholars to collaborate with based on social and academic similarity. The tool uses a Venn diagram together with a tag cloud to explain the similarity between the publications of a user and attendee of a conference. We refer the interested reader to the excellent literature review by \cite{nunes2017systematic} for more examples on explainable RS that provide \textit{Why} explanations.

Compared to providing \textit{Why} explanations, relatively few explainable RS have proposed \textit{How} explanations to provide transparency into the working of the recommendation process. This is mainly due to the fact that in contemporary RS, the underlying algorithm is too complex to be described in a human-interpretable manner (e.g., deep learning models) \cite{zhang2020explainable}. For instance, ‘PeerChooser’ \cite{o2008peerchooser} uses a node-link diagram to explain the recommendation process by highlighting the relationships between user profile attributes (input), social connections, and the recommended items (output). 'SmallWorlds' \cite{gretarsson2010smallworlds} visualizes the inner logic of the recommendation process through a complex network visualization to explain the connection between the active user and the recommended friends. Similarly, ‘TasteWeights’ \cite{bostandjiev2012tasteweights} and ‘LinkedVis’ \cite{bostandjiev2013linkedvis} use a node-link diagram in form of a layer-based interface (three layers) connected via outgoing links to visually explain the connections between the user profile and the recommended items. 

In the literature on explainable recommendation, \textit{Why} and \textit{How} explanations are considered as essential to increase transparency \cite{vig2009tagsplanations,alshammari2019mining,tsai2019explaining,kouki2019personalized,ma2021courseq,o2008peerchooser,bostandjiev2012tasteweights,bostandjiev2013linkedvis,jin2016go,parra2014see}, develop trust \cite{ma2021courseq,o2008peerchooser}, and improve user satisfaction \cite{alshammari2019mining,tsai2019explaining,kouki2019personalized,ma2021courseq,bostandjiev2012tasteweights,bostandjiev2013linkedvis} toward the explainable RS. In summary, much work has been conducted on the generation and provision of \textit{Why} and \textit{How} explanations in RS, separately. Even though \textit{Why} and \textit{How} explanations are provided in many explainable RS, less
attention has been paid to how to systematically design these explanation intelligibility types. Moreover, providing both explanation types at the same time within the same RS, and comparing their impacts on the users' perceptions of transparency, trust, and user satisfaction are under-explored in the literature on explainable recommendation. As pointed out by \citet{lim2009why}, different intelligibility types have different impacts on users' mental models of the system and would result in changes in users' perceptions of trust and user experience. Partly, the reason is that the effectiveness of an
explanation is relative to the question asked \cite{liao2020questioning}. Our work aims to fill that gap. We deal with providing and comparing the value of explanations that address \textit{Why} and \textit{How} intelligibility questions to investigate when and if these explanations are beneficial to users. 
Concretely, we look into how to systematically design and provide both \textit{Why} and \textit{How} explanations side-by-side in the transparent Recommendation and Interest Modeling Application (RIMA) and explore the potential effects of these two explanation intelligibility types on users' perceptions of transparency, trust, and user satisfaction with the application. 
\section{RIMA Application}
To answer our research question, we developed the transparent Recommendation and Interest Modeling Application (RIMA) that offers explainable interest models and recommendations. RIMA is a content-based RS that produces content-based explanations. It offers explanations on demand by allowing users to decide whether to inspect the explanation or not, and empowering them to steer the explanation process as desired. In this work, we focus on recommending scientific publications and use explanatory visualizations to provide \textit{Why} and \textit{How} visual explanations, aiming to clarify the background behavior of the RS by (a) making users aware of the inputs, (b) revealing the system's inner workings, and (c) justifying the recommendation results. The user interest models in RIMA are automatically inferred from users’ publications. The recommendation engine uses these generated interest models to provide scientific publication recommendations. Specifically, the system utilizes the top five interests, based on their respective weights, as the initial input for the recommendation process. The semantic scholar API is employed to retrieve candidate publications that are related to one or more of the user interests. A keyphrase extraction algorithm is then applied to extract keywords from the title and abstract of the fetched publications. Word embedding techniques are utilized to generate vector representations of the user interest model (based on user's top five interests) and the extracted publication (based on the set of keywords generated from it). After that, the cosine similarity between these two embedding representations is calculated to determine the semantic similarity score. The top ten publications with a semantic similarity score exceeding a threshold of 40\% are displayed to the user.
\section{Designing \textit{Why} and \textit{How} Explanations in RIMA}
As an elegant translation of explanations should be carefully designed so that they are satisfying and easy for humans to understand \cite{mohseni2021multidisciplinary}, we decided to use the Human-Centered Design (HCD) approach \cite{norman2013design} and the What-Why-How visualization framework \cite{munzner2014visualization} to systematically design interactive visualizations of the \textit{why} and \textit{How} explanations in RIMA. The HCD approach consists of four consecutive activities, namely \textit{Observation}, \textit{Ideation}, \textit{Prototyping}, and \textit{Testing}. Designing with the HCD ensures that the needs and requirements of the user are taken into consideration as it is based on involving users from the very beginning and regularly consulting them for incremental prototype evaluations. The What-Why-How visualization framework is a high-level framework for analyzing visualizations in terms of three questions: "What" data the user sees (i.e. data), "Why" the user intends to use the visualization (i.e., task), and "How" the visual encoding and interactions are constructed in terms of design choices (i.e., idiom) \cite{munzner2014visualization}. In order to evaluate different prototypes for each explanation, a group of potential users (i.e., researchers and students who were interested in scientific literature) was selected to participate in the design process. For each design iteration, five different users were involved to test and give feedback on the provided prototypes, as recommended by Nielsen \cite{nielsen2000five}. We arrived at the final design of the \textit{Why} and \textit{How} explanations after three HCD iterations, which are described below in detail.

\subsection{First Iteration}
Through this initial step, we aimed at understanding users' needs and requirements to initiate the first low-fidelity prototypes for the \textit{Why} and \textit{How} explanations. 

\subsubsection{Observation:} We started by conducting interviews with five potential users in order to determine the requirements for \textit{Why} and \textit{How} explanations in a scientific literature RS. Through the interviews, we investigated (1) users' expectations from \textit{Why} and \textit{How} explanations, and (2) what kind of visualizations would help them most to understand these explanations. 
The outcomes of the interviews have led to a deeper understanding of the end-users' needs and expectations.

For the \textit{Why} explanation, the interviewees reached a consensus that the explanation should show keywords similar to the user's interests. They suggested that such keywords should be either highlighted using distinctive colors, bold font, or underlining. In terms of the expected content of an explanation, the participants expressed interest in knowing the frequency of their preferred keywords in the recommended publications and proposed the use of a tooltip to display the frequency that can be accessed by hovering over the keywords. They also suggested illustrating the similarity between the user interests and the recommended publication using simple visualizations such as a bar chart or word cloud diagram. 
To simplify the presentation of the information and avoid any confusion by the amount of information, they suggested displaying similar keywords initially, with supplementary details (e.g., frequency) becoming available upon clicking on the keywords (i.e., detail on demand).

Regarding the \textit{How} explanation, the interviewees expected that this explanation should contain information on the system's functionality and the underlying algorithm. They proposed using a simple flow chart depicting how the recommendations are generated using their real data (i.e., interests). They suggested that the flow chart should be interactive, such that additional details would become available upon interaction, as users preferred not to be overwhelmed with too much information at once (i.e., detail on demand). Furthermore, they mentioned that they want to see the algorithm's name and that it should be clickable, allowing them to obtain more information about it.  

\subsubsection{Ideation:} The ideation phase was focused on generating ideas on how to provide interactive visual \textit{Why} and \textit{How} explanations that address the users' preferences and needs identified during the observation phase. A brainstorming session was conducted involving five authors and eight students from the local university with expertise in RS and information visualization to generate as many ideas as possible for each explanation. The primary objective of the brainstorming session was to focus on quantity rather than quality. For both \textit{why} and the \textit{how} explanation, each idea was discussed following a “pitch and critique” approach in order to gather positive and negative feedback. Subsequently, these ideas were analyzed using the What-Why-How visualization framework. As each what-why-how question has a corresponding data-task-idiom answer, every idea for each explanation was presented as a visualization instance defined by its data and task abstraction and the corresponding visual encoding idioms and interaction idioms. The last step was the voting process to select the best ideas. 
After conducting the ideation phase and analyzing the data types and users’ \{action, target\} pairs, a decision was made to present information in a step-by-step manner to enhance user comprehension of explanations. Following the information visualization rules of thumb proposed in \cite{munzner2014visualization}, for both \textit{Why} and \textit{How} explanations we start with an overview first, then through interaction mechanisms we provide details on demand. This approach was chosen to prevent overwhelming the users with an excessive amount of information and minimize confusion. 

For the \textit{Why} explanation, the data to be shown include the user's interests (categorical data), the keywords extracted from the publications' title and abstract (categorical data), and the similarity score between them (quantitative data). 
The possible tasks, expressed as \{action, target\} pairs, that an idiom for \textit{Why} explanation could provide are \{Present, Similarity\}, \{Discover, Dependency\}, and \{Summarize, Dependency\} between the recommended publication and the user's interests. 
Depending on the data to be shown and the aimed task, potential idioms were selected. Starting by providing an overview, visual elements such as coloring and highlighting were used to indicate the relevance between a recommended publication and the user's interests. A similarity score is also provided, which can be either between all the keywords and the interests or a particular keyword and its related interests. The former similarity score is displayed next to the recommended publication, while the latter will appear by hovering over a particular keyword, where a pop-up will show up displaying the similarity score between that keyword and its related interests. More details about the reason for recommending a specific publication are provided on demand through a bar chart and word cloud as idioms.
Interactions provided in these visualizations allow users to manipulate the view by selecting elements (e.g., keywords) from within the view, facet data between views by juxtaposing and coordinating multiple views (e.g., bar chart and word cloud), and reduce the data within a single view by slicing the data attributes and showing only items that match a specific value for the given attribute.

In the \textit{How} explanation, the data to be shown is the user's interests (categorical data), the keywords extracted from the publications' title and abstract (categorical data), the similarity score between the two (quantitative data), and vector representations of both interests and keywords (quantitative data). 
The possible tasks expressed as \{action, target\} pairs) that an idiom for \textit{Why} explanation could provide are
The tasks that an idiom for \textit{How} explanation could provide include \{Discover, Dependency\}, \{Present, path\}, and \{Summarize, path\} of the process of generating the recommendation outcomes. 
Depending on the data to be shown and aimed task, the flow chart idiom was selected to visualize the inner working of the system. Flow charts have been found to be an effective tool for illustrating complex processes for users of all skill levels, increasing their engagement with and understanding of a topic \cite{jin2016go}. 
The provided interactions in these visualizations enable users to manipulate the visualization by navigating within the view using a navigation panel to facilitate the movement from an overview explanation to a more detailed one, and facet data between different views by partitioning data between them (e.g., set of interests and set of keywords).
\subsubsection{Prototyping:} 
Following the analysis of user requirements obtained from interviews and the outcome of our brainstorming session, a series of low-fidelity prototypes were created using a digital pen and tablet. We asked the users to select the most appropriate prototypes in terms of content (i.e., information) and display (i.e., visualization).
Low-fidelity prototypes are considered to be essential due to their simplicity and quick creation process, which enable non-designers to effectively communicate their ideas in real time.

\begin{figure}[h]
     \centering
     \begin{subfigure}[b]{0.49\textwidth}
         \centering
         \includegraphics[width=0.9\textwidth, height=0.6\textwidth]{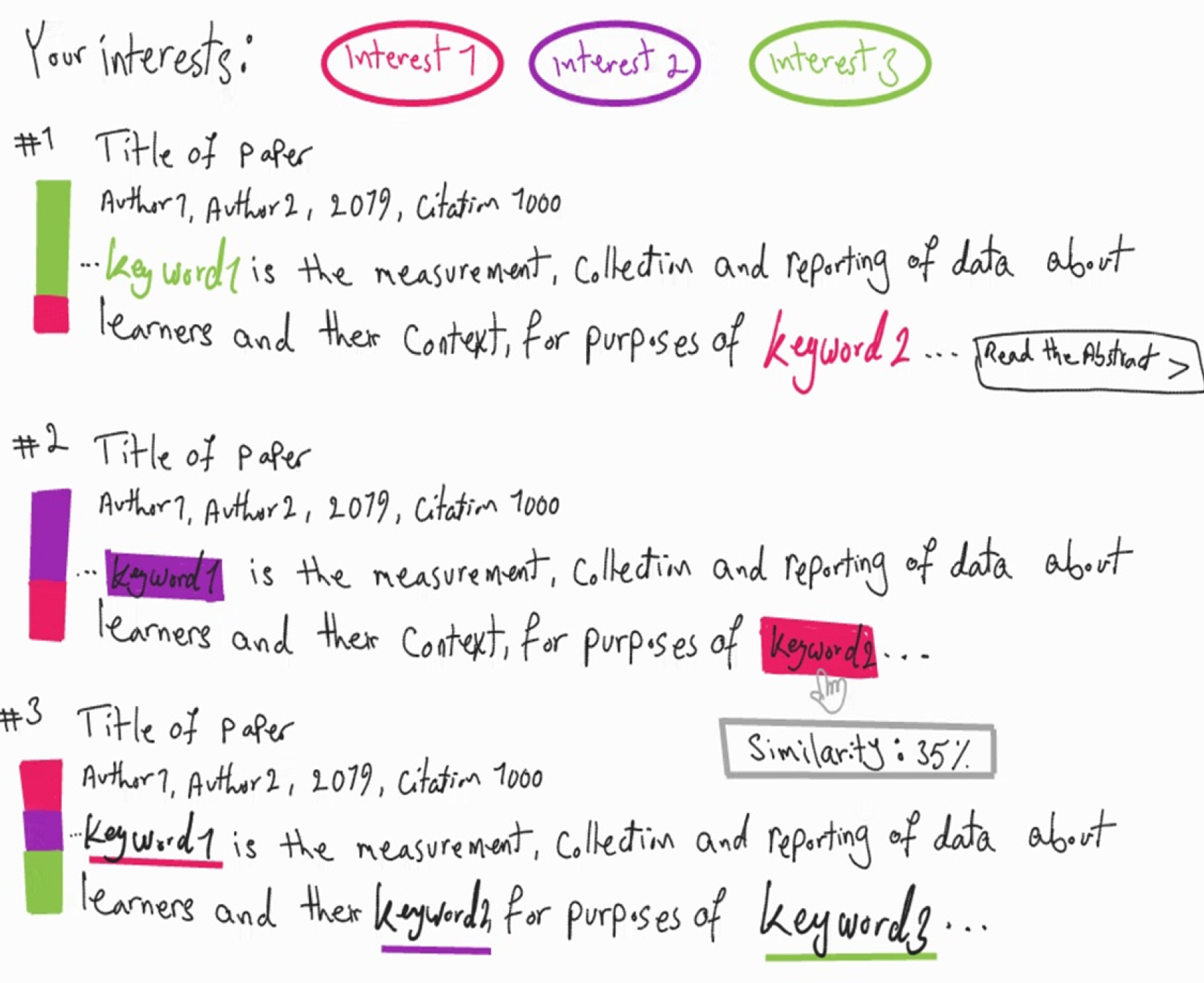}
         \caption{Overview}
         \label{fig:why_overview_I1}
     \end{subfigure}
     \hfill
     \begin{subfigure}[b]{0.49\textwidth}
         \centering
         \includegraphics[width=0.9\textwidth, height=0.6\textwidth]{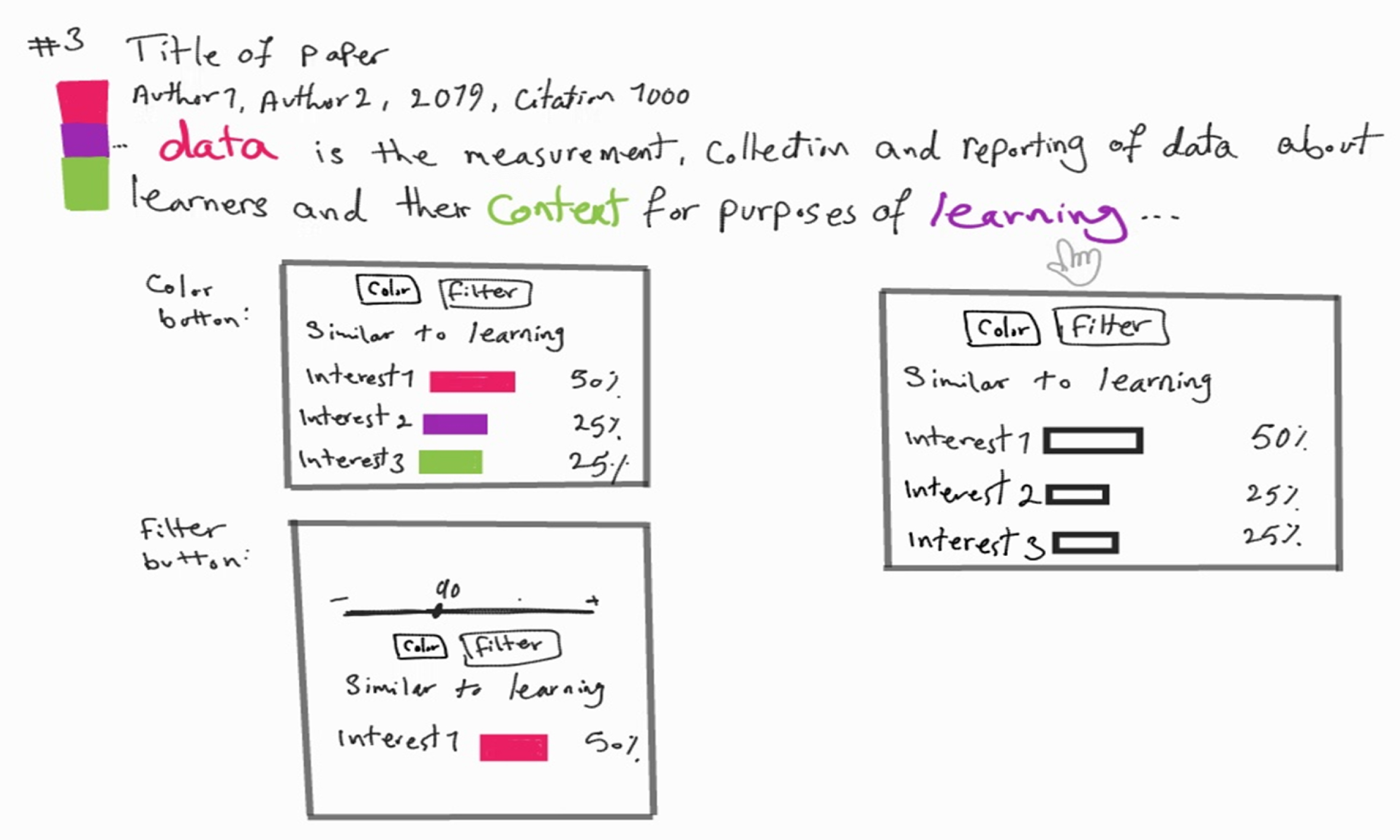}
         \caption{Detailed}
         \label{fig:why_detailed_I1}
     \end{subfigure}
        \caption{ \textit{Why} explanation – Iteration 1} 
        \label{fig:Why_I1}
\end{figure}

For the \textit{Why} explanation low-fidelity prototype, users leaned toward the idea of starting with "overview first" by presenting an overview through the use of color bands next to each publication, with the same colors used for the user interests at the top, where the height of each color band reflects the relevance of the publication to the related interest (Figure \ref{fig:why_overview_I1}). The user can interact with this view by hovering over the keywords, where a tooltip will indicate the interests to which a keyword is similar based on their similarity scores. 
To provide a more detailed explanation, a "why" button is included, leading to more detailed visualizations, including a word cloud as an initial view containing all the keywords extracted from the publication abstract, and a bar chart that appears upon hovering over a keyword from the initial view illustrating the similarity between the selected keyword and the corresponding relevant user's interests (Figure \ref{fig:why_detailed_I1}).

\begin{figure}[h]
     \centering
     \begin{subfigure}[b]{0.35\textwidth}
         \centering
         \includegraphics[width=0.95\textwidth, height=0.7\textwidth]{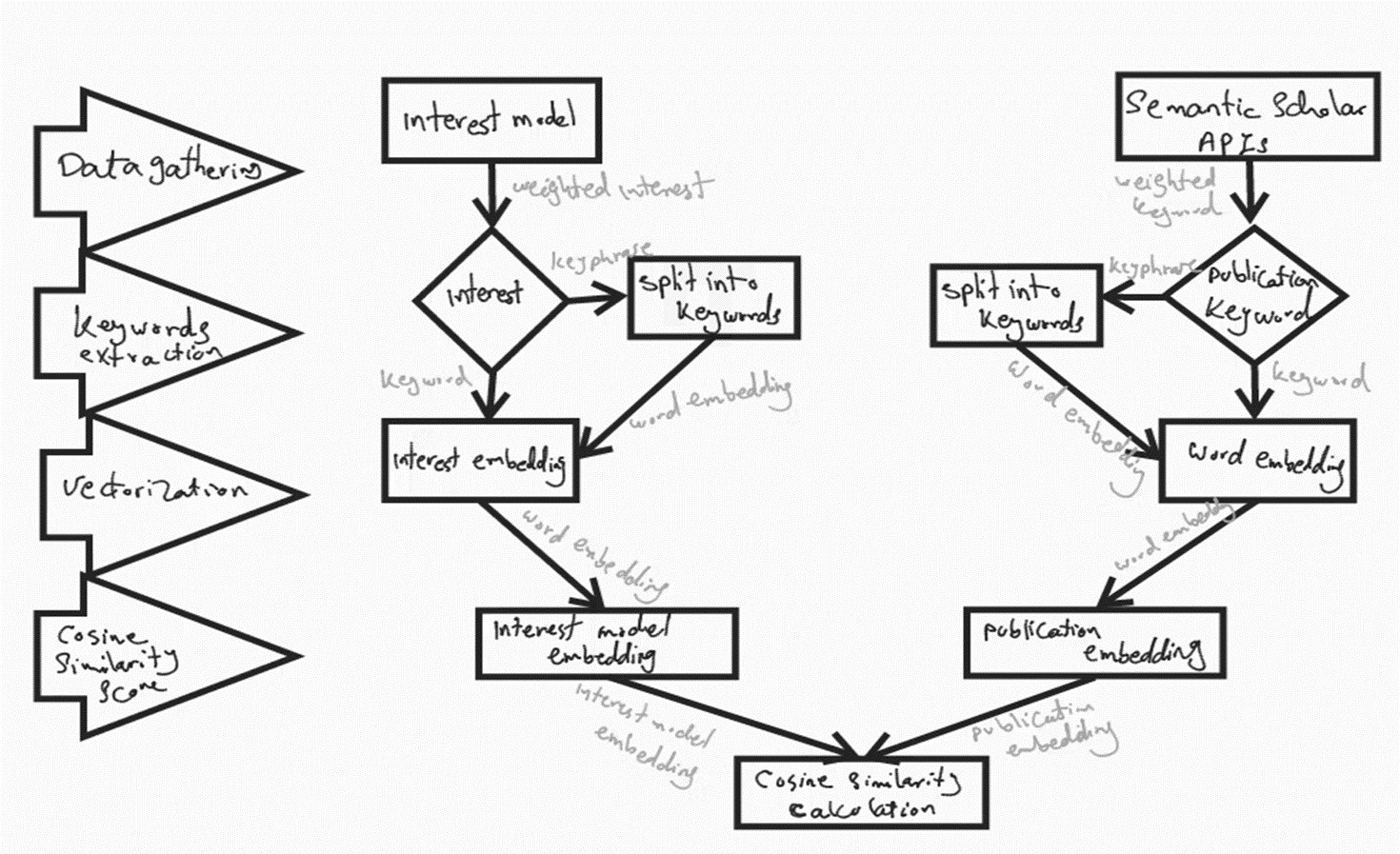}
         \caption{Overview}
         \label{fig:how_overview_I1}
     \end{subfigure}
     \hfill
     \begin{subfigure}[b]{0.6\textwidth}
         \centering
         \includegraphics[width=\textwidth, height=0.4\textwidth]{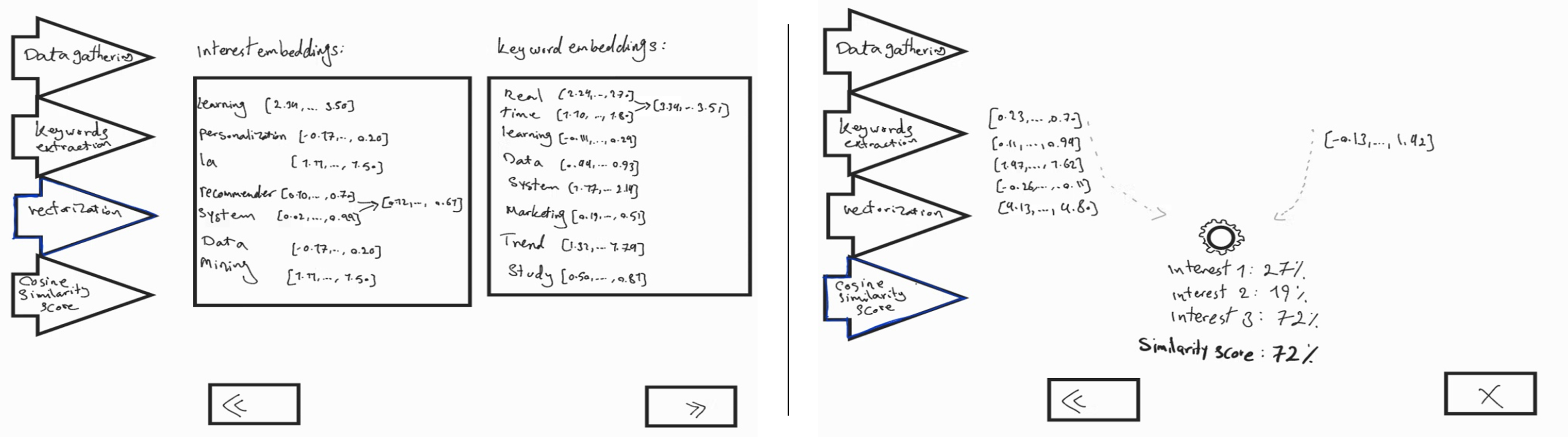}
         \caption{Detailed}
         \label{fig:how_detailed_I1}
     \end{subfigure}
        \caption{ \textit{How} explanation – Iteration 1} 
        \label{fig:How_I1}
\end{figure}

Likewise, users preferred beginning the \textit{How} explanation with an "overview first" and then providing "detail on demand".
The user can access this explanation from the \textit{Why} explanation view by clicking on a "how" button. Initially, the \textit{How} explanation provides an overview of the inner working of the system by explaining how recommendations are generated using a flow chart illustrating the main steps of the algorithm (Figure \ref{fig:how_overview_I1}). In addition, a navigation panel is available to the user, which displays the main steps of the algorithm. By clicking on each step, the user can access more information and adjust their view accordingly. 
Similarly, the flow chart idiom is selected for visualizing each step in more detail (Figure \ref{fig:how_detailed_I1}). 

\subsubsection{Testing:}
The objective of evaluating the initial low-fidelity prototypes is twofold: First, to obtain feedback that can be used to improve and optimize the design, and second, to understand to what extent each of the selected visualizations was able to convey the intended purposes of the explanations to the user. This feedback was collected through a qualitative evaluation with the same five users who participated in the observation phase, using a think-aloud approach. Furthermore, we used open-ended questions to ask users about their thoughts on each of the selected visualizations and their opinion towards the proposed initial low-fidelity prototypes for the \textit{Why} and \textit{How} explanations. 

For the \textit{Why} explanation, the users demonstrated a preference for the coloring feature over the keyword highlighting feature and to present keywords in a bold font in addition to being colored. Moreover, they suggested using the same colors in the word cloud idiom and having the legends outside the bars in the bar chart idiom. Furthermore, the users expressed a desire to compare a specific keyword with each relevant interest instead of one similarity score provided. Overall, all users agreed on the selected idioms and felt satisfied with the explanation content. 
Regarding the \textit{How} explanation, the users expressed disagreement with the labels in the navigation panel and suggested using a clearer naming. Moreover, they proposed simplifying the overview flow 
chart by presenting it with fewer levels of information.
\subsection{Second Iteration}
At this stage, we aimed to address the shortcomings of the previous designs by considering users’ feedback collected from the testing phase. In this iteration, prototypes are designed using the Figma tool, but they are still considered to be low-fidelity.

\subsubsection{Prototyping:}
\begin{figure}[h]
     \centering
     \begin{subfigure}[b]{0.49\textwidth}
         \centering
         \includegraphics[width=\textwidth]{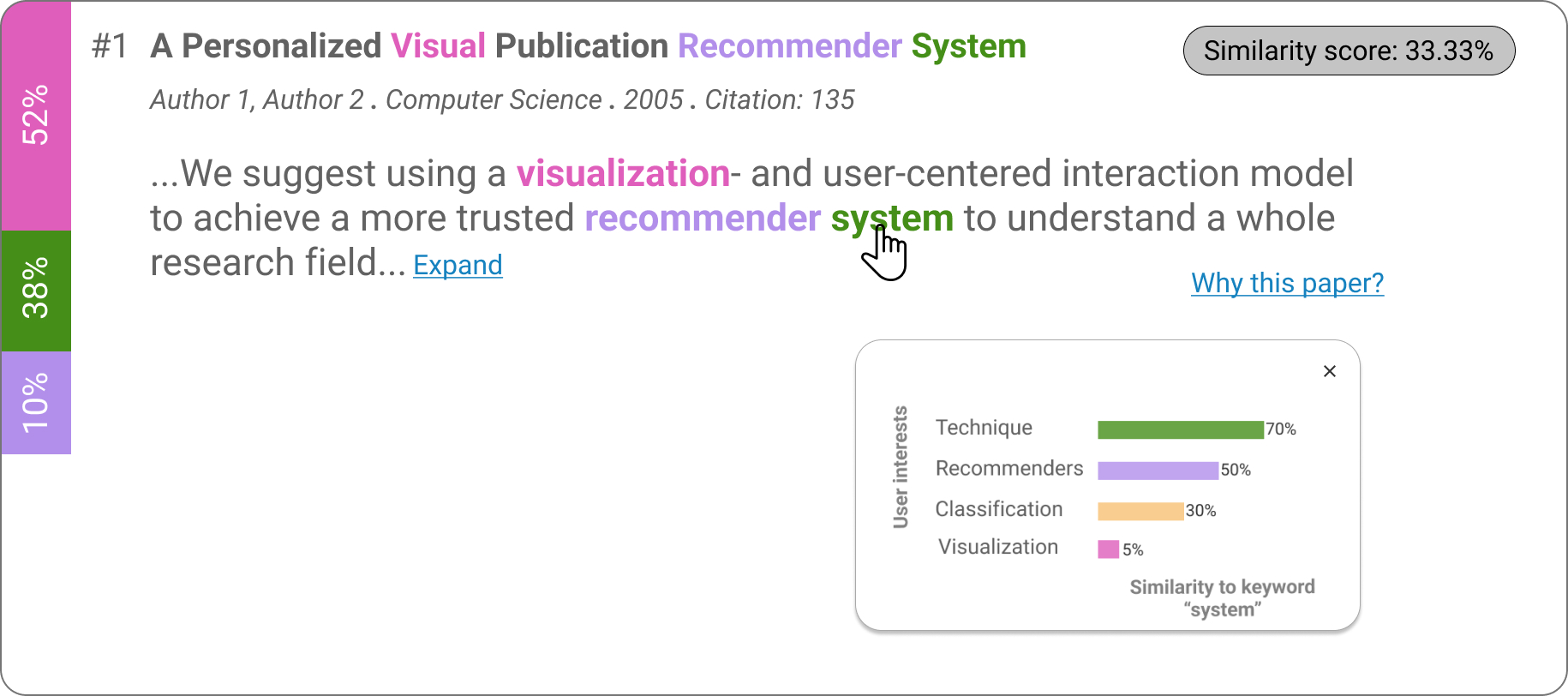}
         \caption{Overview}
         \label{fig:why_overview_I2}
     \end{subfigure}
     \hfill
     \begin{subfigure}[b]{0.49\textwidth}
         \centering
         \includegraphics[width=\textwidth]{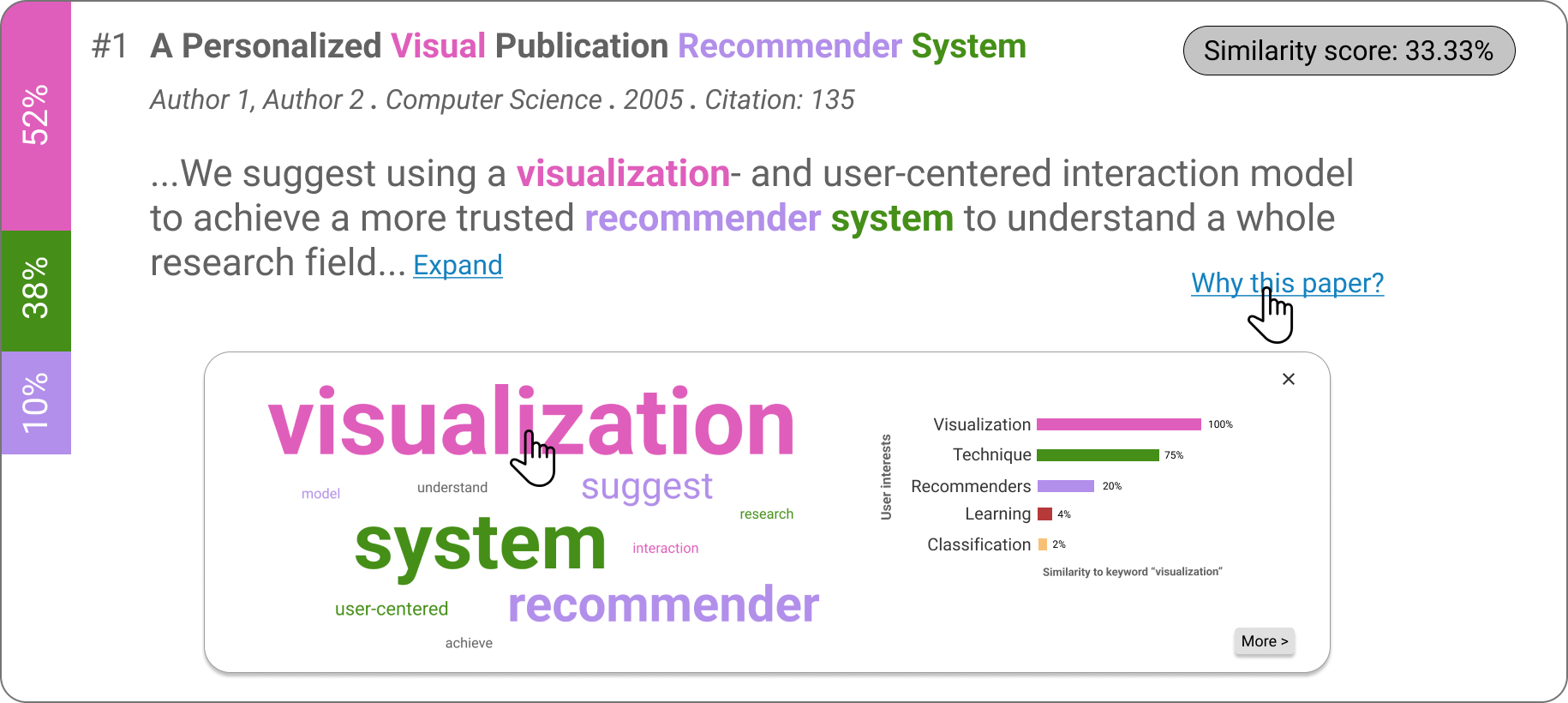}
         \caption{Detailed}
         \label{fig:why_detailed_I2}
     \end{subfigure}
        \caption{ \textit{Why} explanation – Iteration 2} 
        \label{fig:Why_I2}
\end{figure}

In the \textit{Why} explanation, the keywords are presented in bold and colored in the initial view (Figure \ref{fig:why_overview_I2}). Additionally, the similarity score previously shown when hovering over a keyword in the abstract has been replaced with a bar chart that allows users to easily visualize the degree of similarity between a given keyword and all relevant interests. Upon clicking on the "Why this paper?" button, users are directed to a second visualization where colored keywords are displayed within a word cloud (Figure \ref{fig:why_detailed_I2}). By hovering over any given keyword, users can view a corresponding bar chart that illustrates the degree of similarity between the selected keyword and all five user’s interests. 

\begin{figure}[h]
    \centering
         \includegraphics[width=0.95\textwidth]{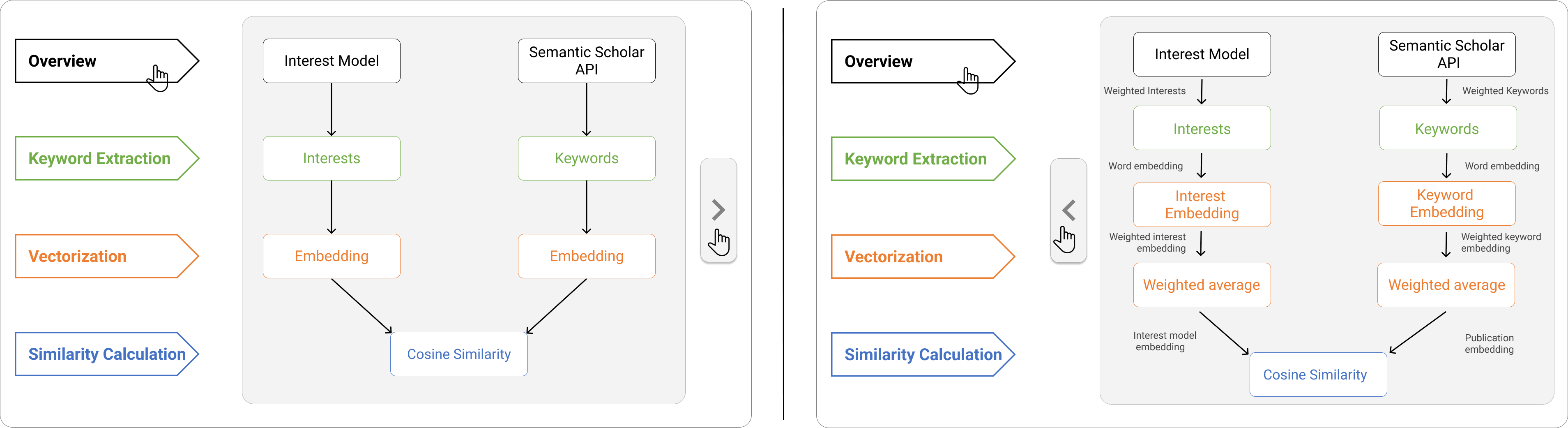}
         \caption{\textit{How} explanation Overview – Iteration 2}
         \label{fig:how_overview_I2}
\end{figure}

\begin{figure}[h]
    \centering
         \includegraphics[width=0.6\textwidth, height=0.95\textwidth]{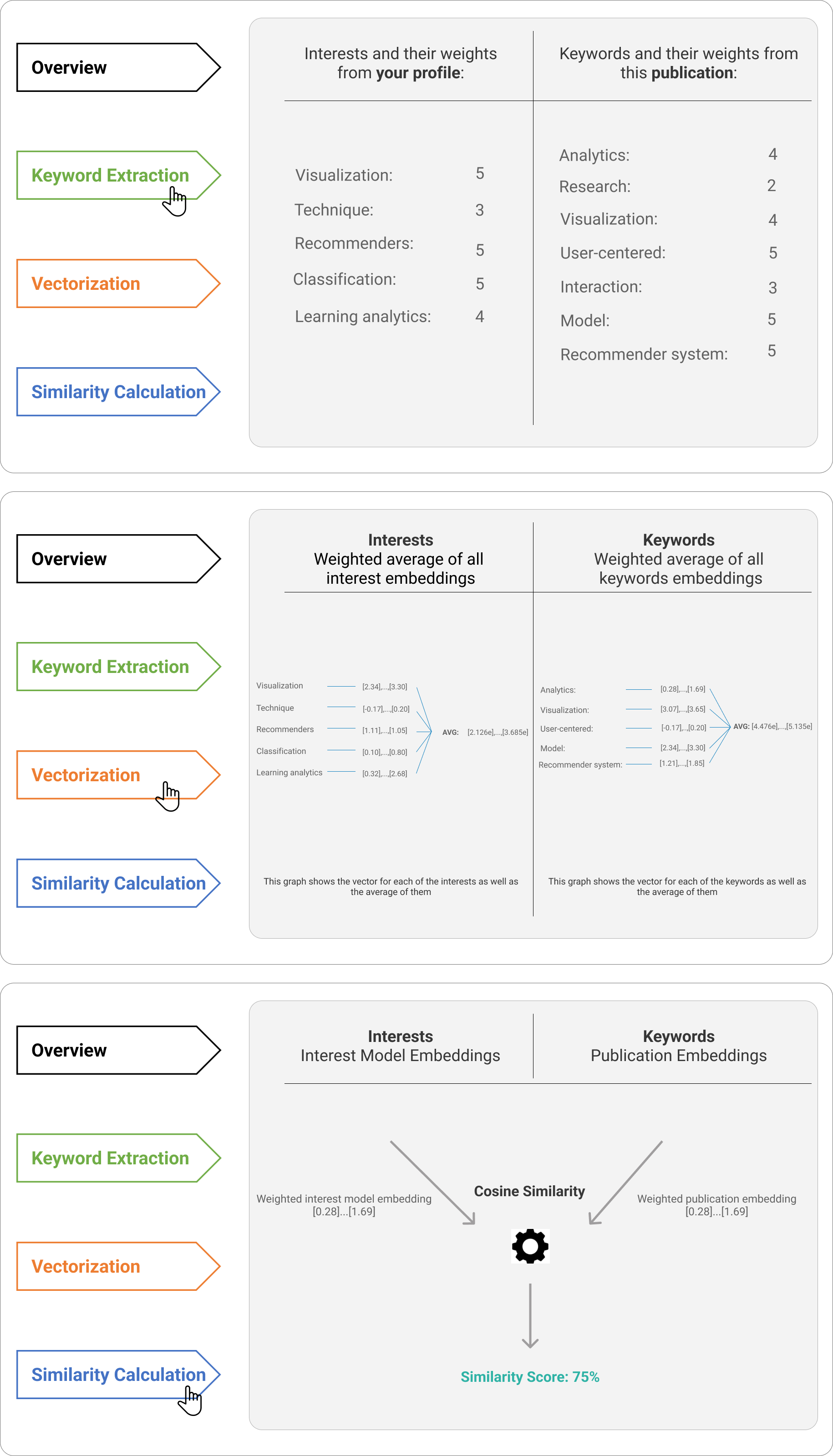}
         \caption{\textit{How} explanation Detailed – Iteration 2}
         \label{fig:how_detailed_I2}
\end{figure}

Regarding the \textit{How} explanation, as suggested by users, we initially provide a flow chart with two levels of information detail illustrating an overview of the main steps of the inner working of the system and explaining how recommendations are generated. Users can toggle between the two levels of the flow chart using an arrow button (Figure \ref{fig:how_overview_I2}). For the detailed view, users can navigate through the three main steps (i.e., Keyword Extraction, Vectorization, Similarity Calculation) of the underlying algorithm using the left navigation panel where we updated the labels of the buttons in order to make them more understandable (Figure \ref{fig:how_detailed_I2}). 

\subsubsection{Testing:}
In the second evaluation round, five new users were asked to provide feedback to the Figma prototypes. For the \textit{why} explanation, users were satisfied with the provided visualizations and they reported that it helped them understand the reason behind getting a certain recommendation. Nevertheless, for the bar chart appearing after hovering over a keyword in the abstract, they suggested displaying only three interests with the highest similarities to the selected keyword instead of all interests in order to avoid showing low similarities. 

As for the \textit{How} explanation, users mentioned that the labels used are still confusing even after changing them.
Additionally, users found the information provided in different steps unclear and suggested adding a description for each step of the algorithm to make the process easier to understand. They also preferred to keep using the same color for interests and keywords in the flow charts.

\subsection{Third Iteration}
\begin{figure}[h]
     \centering
     \begin{subfigure}[b]{0.49\textwidth}
         \centering
         \includegraphics[width=\textwidth]{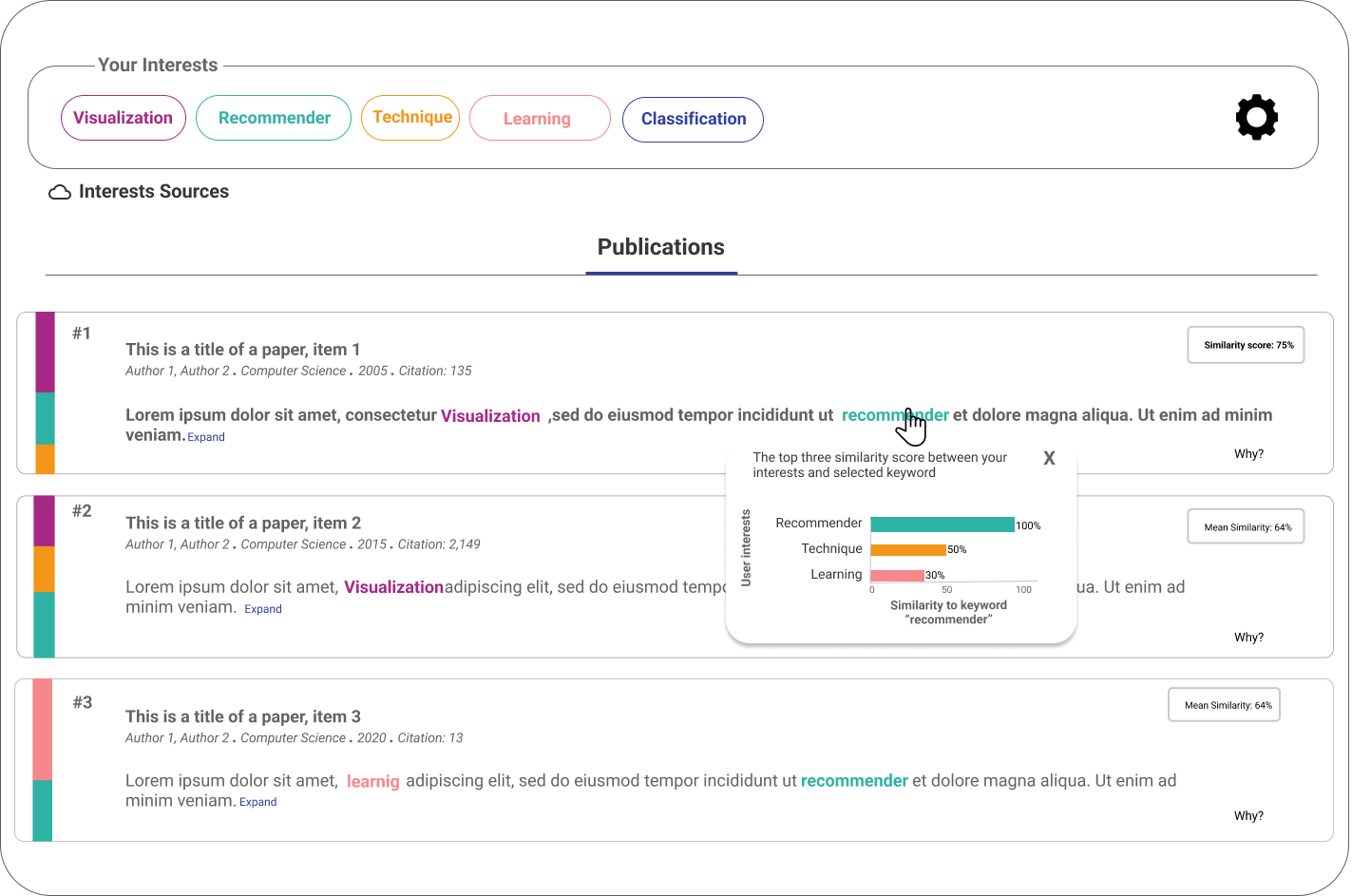}
         \caption{Overview}
         \label{fig:why_overview_I3}
     \end{subfigure}
     \hfill
     \begin{subfigure}[b]{0.49\textwidth}
         \centering
         \includegraphics[width=\textwidth]{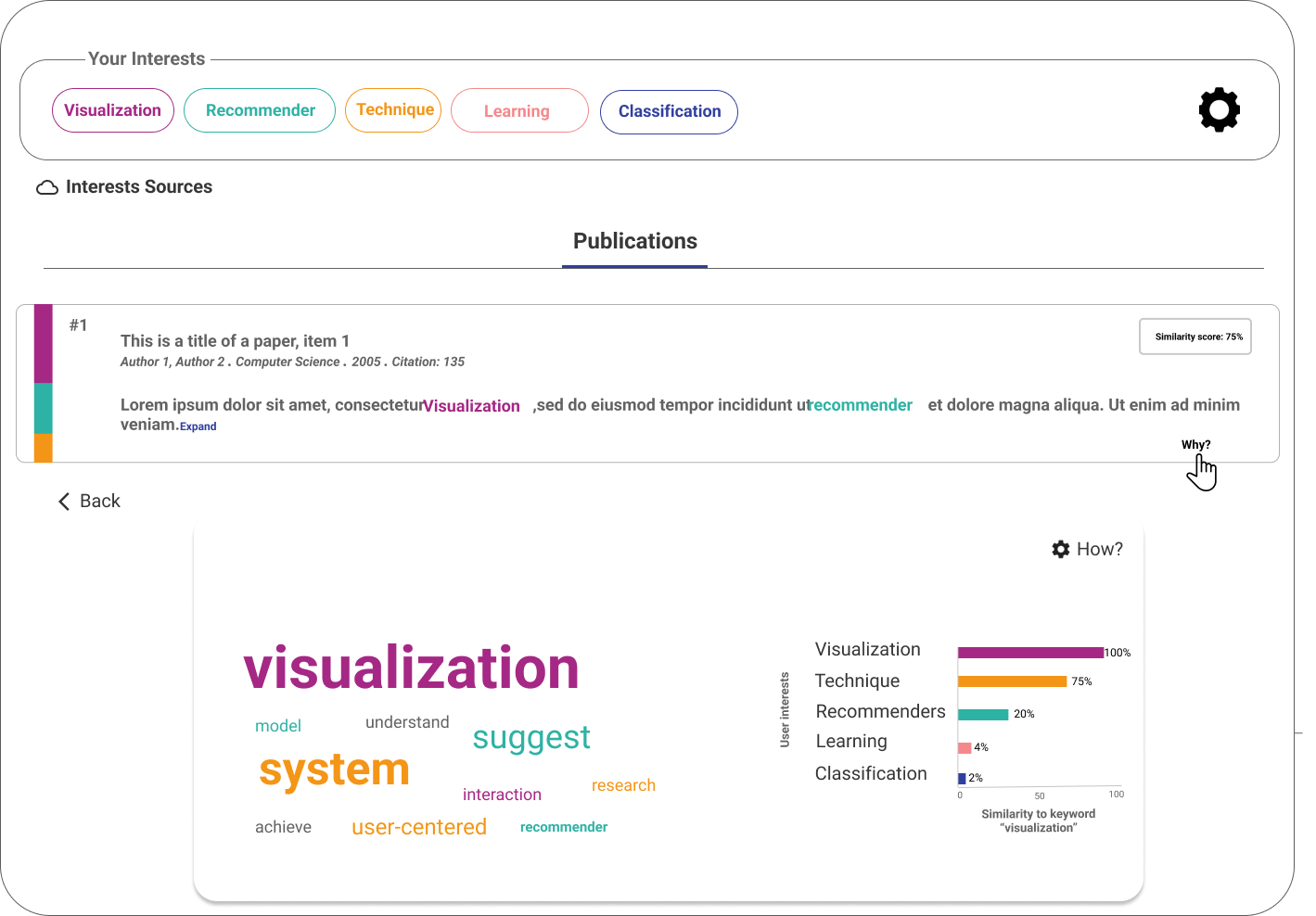}
         \caption{Detailed}
         \label{fig:why_detailed_I3}
     \end{subfigure}
        \caption{ \textit{Why} explanation – Iteration 3} 
        \label{fig:Why_I3}
\end{figure}

\begin{figure}[h]
    \centering
         \includegraphics[width=0.95\textwidth]{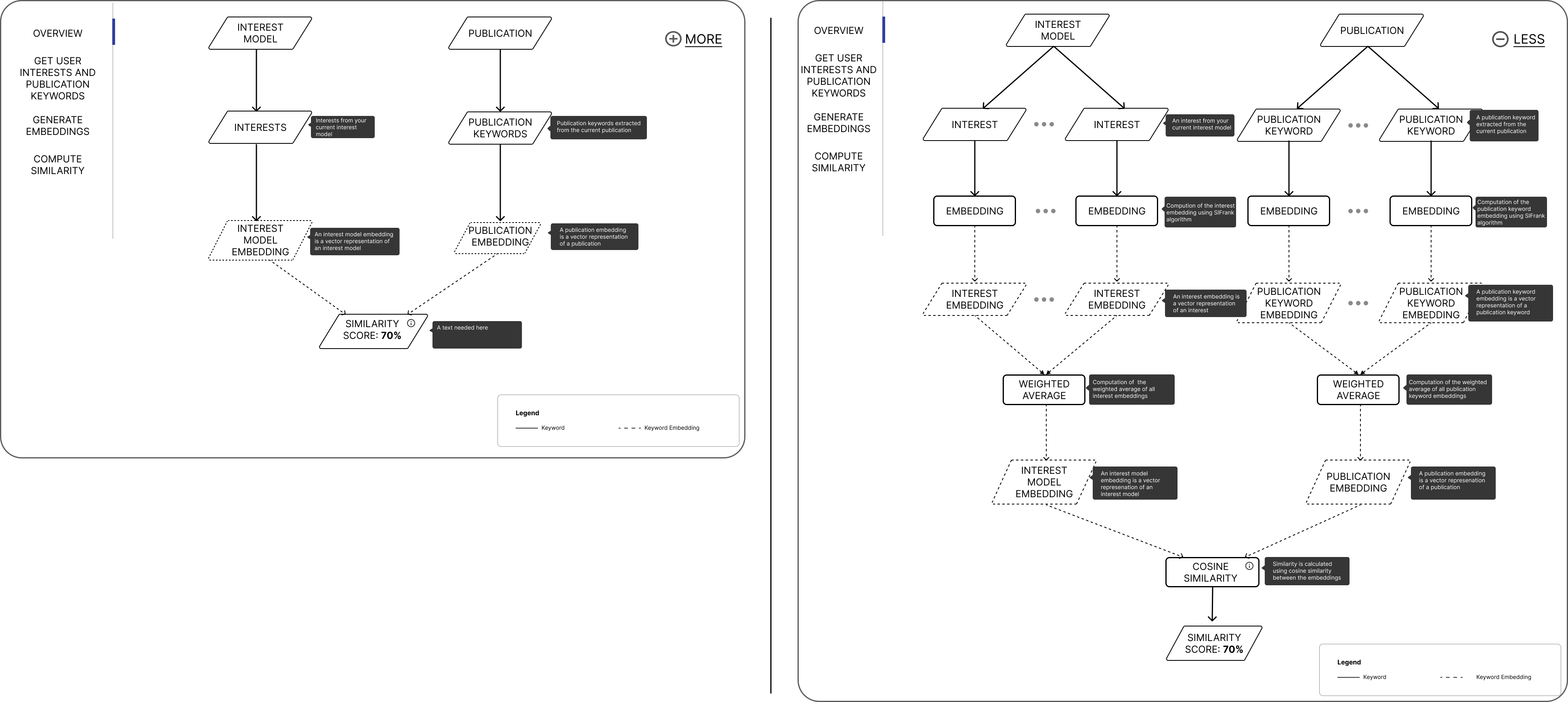}
         \caption{\textit{How} explanation Overview – Iteration 3}
         \label{fig:how_overview_I3}
\end{figure}

\begin{figure}[h]
    \centering
         \includegraphics[width=0.65\textwidth, height=0.99\textwidth]{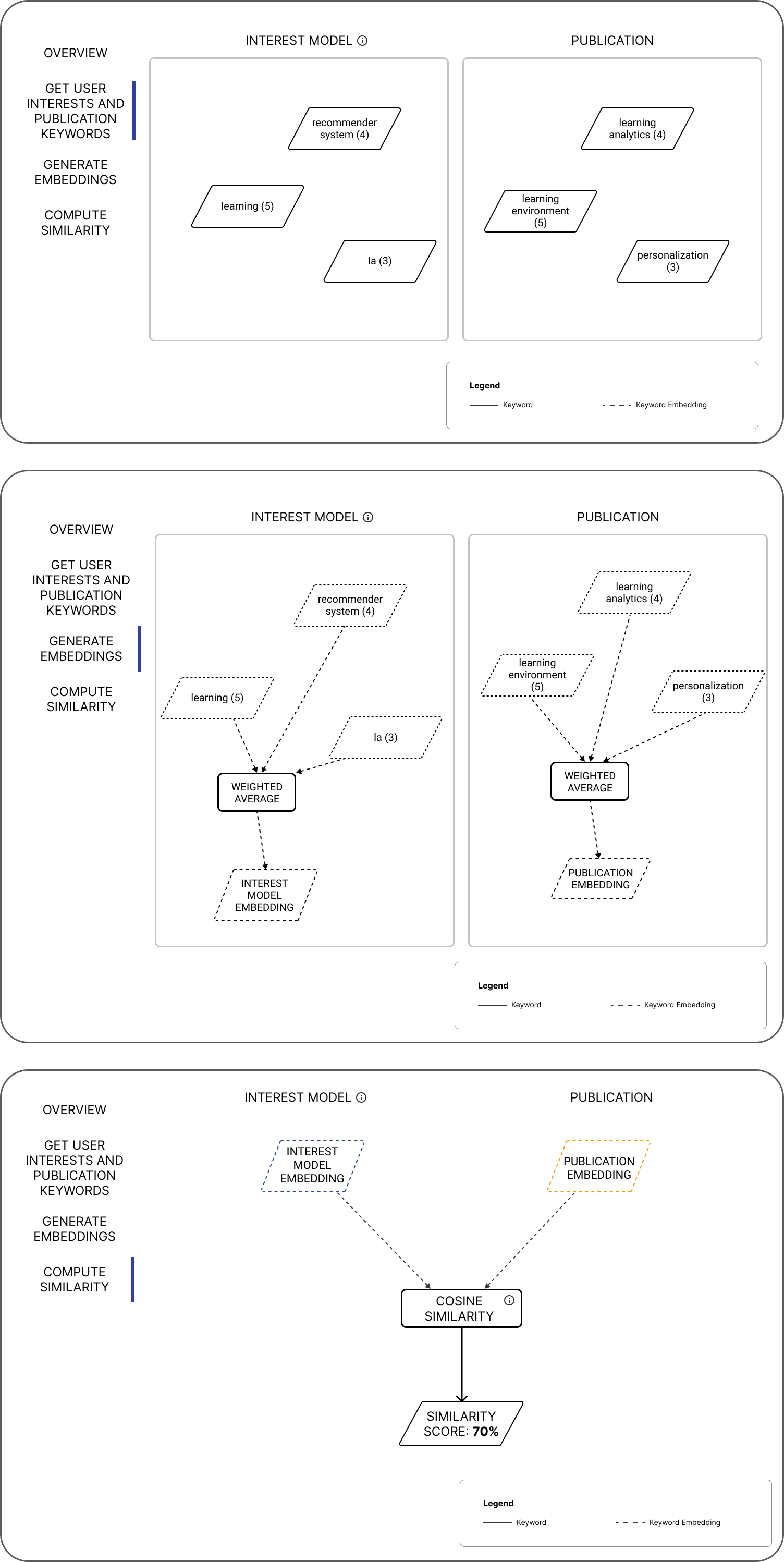}
         \caption{\textit{How} explanation Detailed – Iteration 3}
         \label{fig:how_detailed_I3}
\end{figure}

After incorporating users’ feedback from the second iteration, we proceeded to develop high-fidelity prototypes for the \textit{Why} and \textit{How} explanations using Figma (Figures \ref{fig:Why_I3}, \ref{fig:how_overview_I3}, \ref{fig:how_detailed_I3}). A description of each node has been added in form of a tooltip when hovering over the nodes in the flow chart in the \textit{How} explanation. We then repeated the evaluation process with five new users who were now able to interact with the explanations. We asked the users to think aloud while interacting with the prototypes to gain in-depth feedback. We made minor changes to the prototypes in response to user feedback. Overall, the prototypes were understandable and well received by the users.

\subsection{Implementation}
\begin{figure}[h]
     \centering
     \begin{subfigure}[b]{0.56\textwidth}
         \centering
         \includegraphics[width=\textwidth]{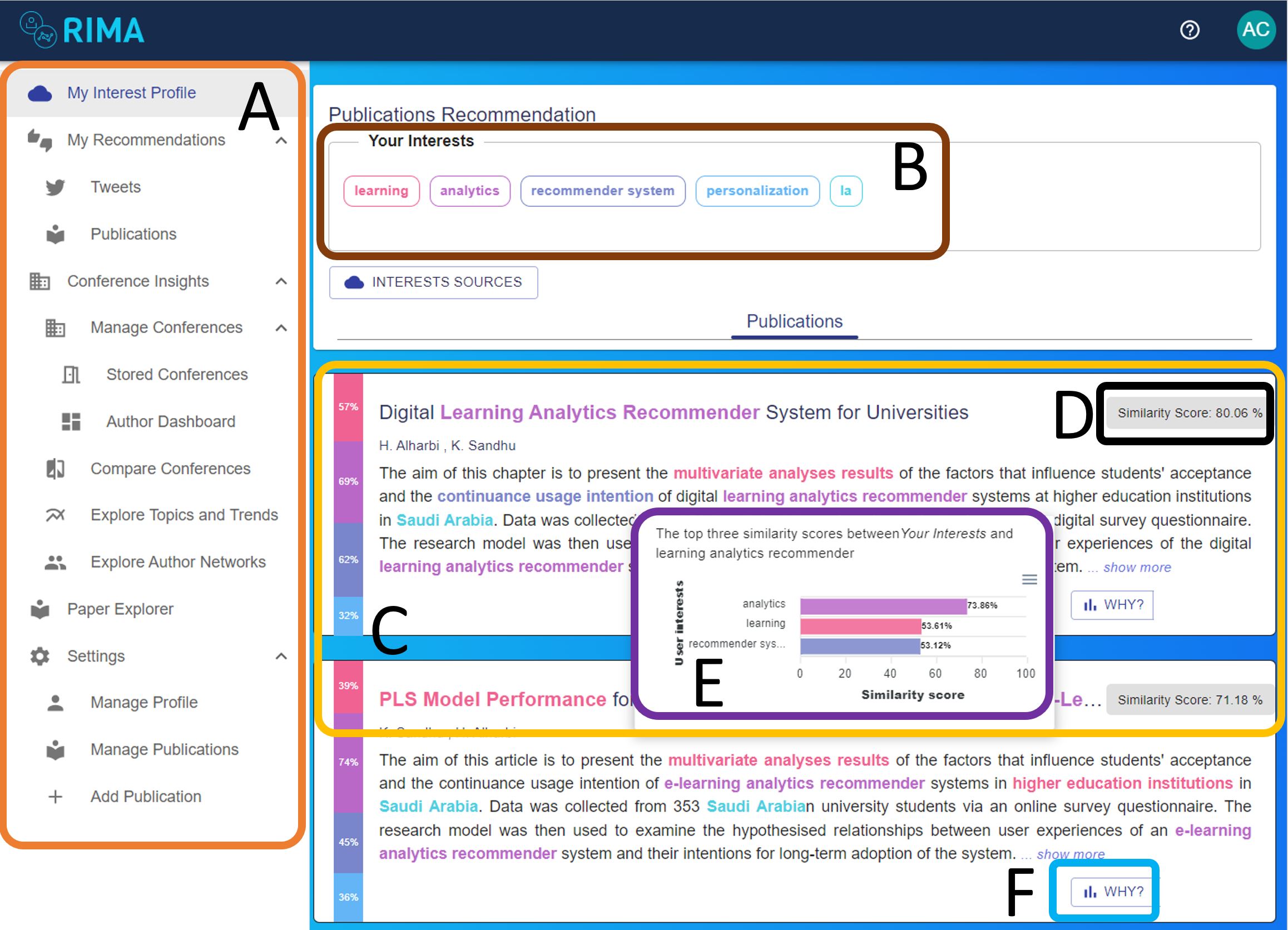}
         \caption{Overview}
         \label{fig:why_overview_imp}
     \end{subfigure}
     \hfill
     \begin{subfigure}[b]{0.43\textwidth}
         \centering
         \includegraphics[width=\textwidth]{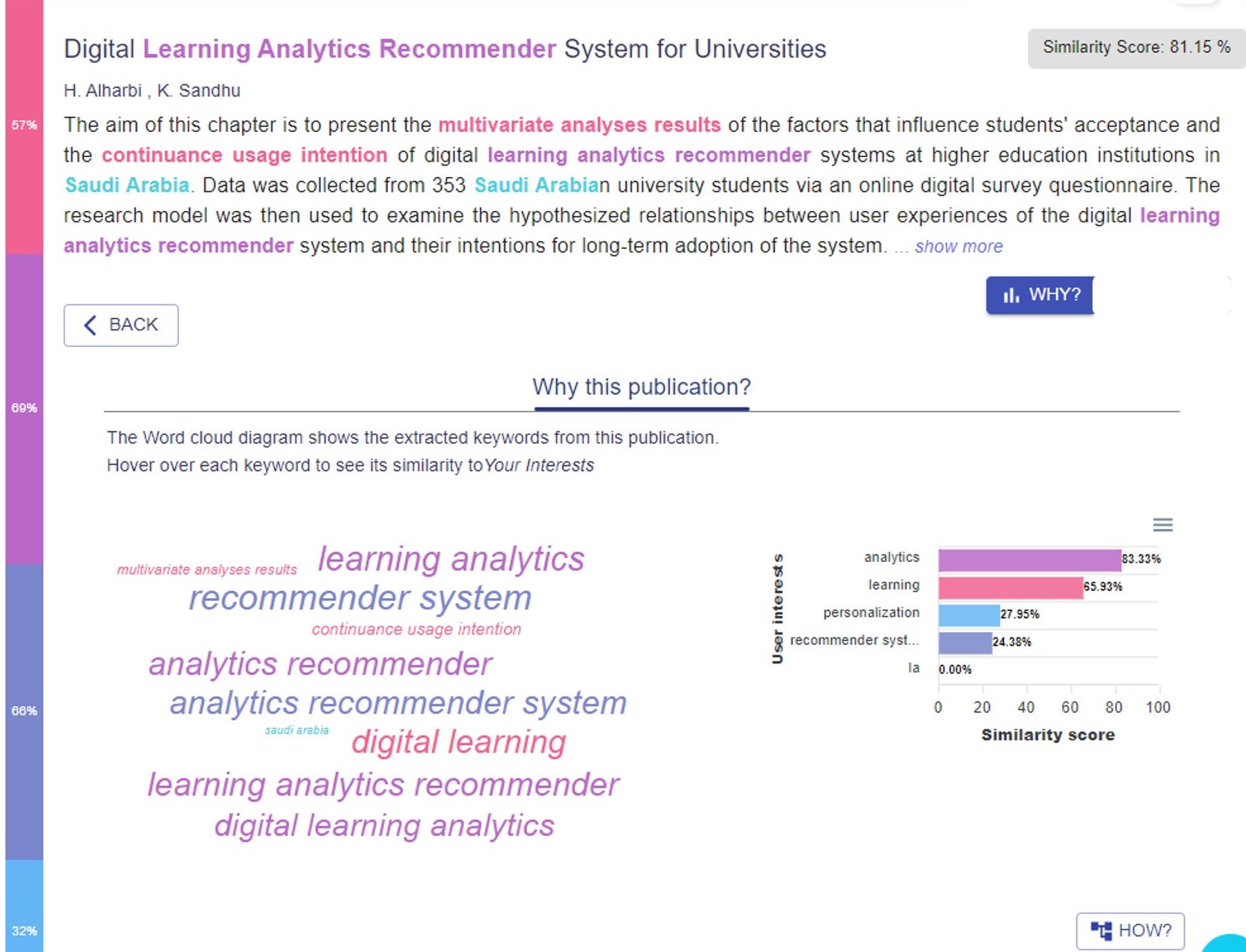}
         \caption{Detailed}
         \label{fig:why_detailed_imp}
     \end{subfigure}
        \caption{ \textit{Why} explanation – Implementation} 
        \label{fig:Why_Imp}
\end{figure}

We implemented the final prototypes of the \textit{Why} and \textit{How} visual explanations in the RIMA application after incorporating users' feedback from previous design iterations. The main interface consists of a navigation panel where the user can access all RIMA services including the explainable recommendation of publications (Figure \ref{fig:why_overview_imp}-A); a list of the top five user’s interests generated automatically by the system with a unique color for each interest to easily identify the interests and their impact on the recommended publications (Figure \ref{fig:why_overview_imp}–B); and a list of the recommended publications in the form of separate boxes (Figure \ref{fig:why_overview_imp}–C). 


In the \textit{Why} explanation, 
we provide an overview explanation using color bands within each recommendation to show the relevance of recommended publications to user's interests, where the height of each color band indicates how relevant is this publication to the related interest. Additionally, we provide a relevance score for each publication (Figure \ref{fig:why_overview_imp}-D). For each recommended publication, a set of keywords are extracted and highlighted in the abstract. Users can interact with these keywords by either hovering over them to see a similarity score to the user interest model or by clicking on them so that a bar chart is shown in a pop-up displaying similarity scores between the clicked keyword and the top three similar interests (Figure \ref{fig:why_overview_imp}-E). 
In order to provide more details, we included a "WHY" button in these boxes (Figure \ref{fig:why_overview_imp}-F) which lead to a more detailed \textit{Why} explanation (Figure \ref{fig:why_detailed_imp}). In this visualization, we provide the information in two steps. Firstly, users are presented with a word cloud that displays the extracted keywords from the current publication. The sizes of these keywords reflect their similarity scores with the user interest model, and they are color-coded to correspond with the  most similar user's interest for each keyword. Secondly, by hovering over each keyword, a bar chart will appear, which depicts the similarity scores between the keyword and all five user's interests.

\begin{figure}[h]
    \centering
         \includegraphics[width=0.95\textwidth, height=0.3\textwidth]{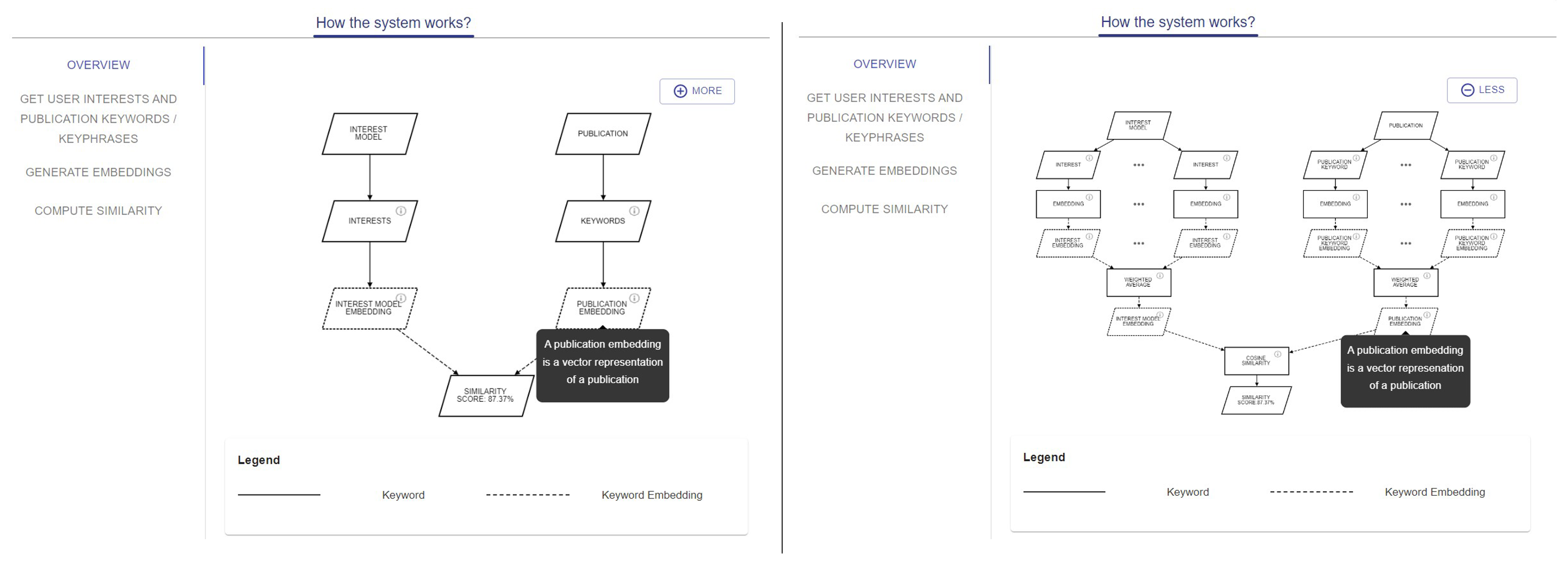}
         \caption{\textit{How} explanation Overview – Implementation}
         \label{fig:how_overview_imp}
\end{figure}

In the \textit{How} explanation, users can learn more about the inner working of the system and how the recommendations are generated by clicking on the "HOW" button in the \textit{Why} explanation interface. Following the method of "overview first, details on demand", we start by presenting an abstract overview of the algorithm, followed by a detailed breakdown of each step using user's actual data (i.e., interests). A flow chart is used to illustrate the process, with buttons and components arranged in a top-down flow for ease of comprehension.
Starting with the "overview first", we offer the \textit{How} explanation with two levels of information detail, accessible via "MORE" and "LESS" buttons, to provide an abstract description of the system's processes (Figure \ref{fig:how_overview_imp}).

\begin{figure}[h]
    \centering
         \includegraphics[width=0.45\textwidth, height=0.99\textwidth]{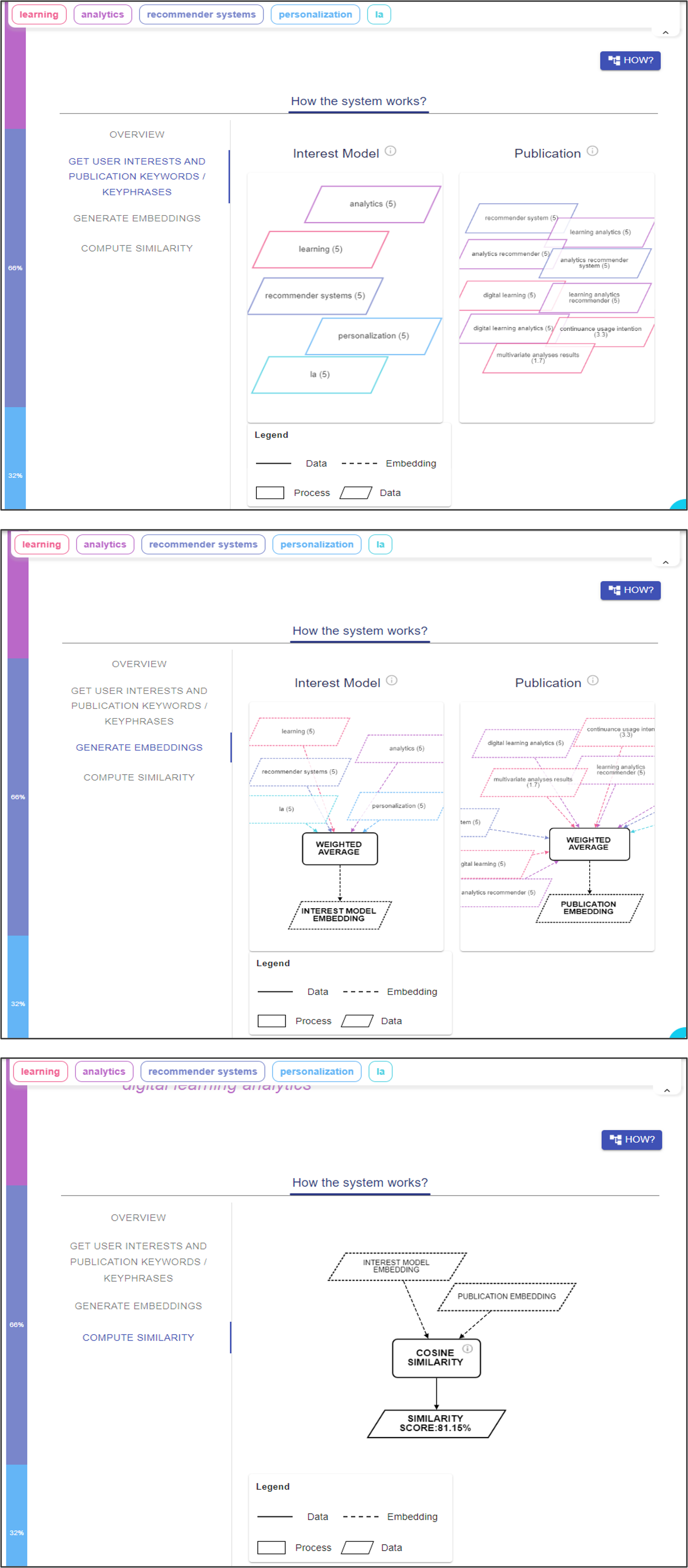}
         \caption{\textit{How} explanation Detailed – Implementation}
         \label{fig:how_detailed_imp}
\end{figure}

In addition to the abstract overview, the left navigation panel displays three distinct steps that illustrate the inner working of the RS algorithm with more technical details, using actual user and publication data (Figure \ref{fig:how_detailed_imp}). Users are able to interact with each step through hovering, zooming, dragging, and dropping. Furthermore, brief descriptions of each node can be viewed by hovering over them. The first step, labeled "Get user interests and publication keywords/keyphrases", displays a visualization of the actual user's interests and the extracted keywords from the publication along with their corresponding weights. The second step, labeled "Generate embeddings", depicts the process of creating vector representations (i.e., embeddings) of the user interest model and the publication. The final step, labeled "Compute similarity", shows how similarity scores are calculated between the interest model and the publication embeddings.
\section{Evaluation}
After systematically designing the \textit{Why} and \textit{How} explanations and implementing them in the RIMA application, we conducted a qualitative user study to explore the usage and attitudes towards our scientific literature RS, considering the \textit{Why} and \textit{How} explanations. We believe that following a qualitative approach is beneficial to investigate in-depth the
users’ unique perspectives and expectations from an explainable RS providing \textit{Why} and \textit{How} explanations together.
\subsection{Study Design}
Researchers and students interested in  scientific literature were invited to participate. 
In total, 12 participants (five females) agreed to take part in this study. Participants were between 20 and 39 years old, where half of them were master’s graduates or higher, and the other half were master’s students. The study included participants from different countries (Germany, Iran, India, Palestine) and study fields (Computer Science, Statistics, Chemical Engineering). All participants gave informed consent to study participation. Participants were initially given a short introductory video about the RIMA application in general, and another short demo video about the \textit{Why} and \textit{How} explanation features in the application. Next, they answered a questionnaire in SoSci Survey which included questions about their demographics and familiarity with RS and visualization. Afterwards, we conducted moderated think-aloud sessions where participants were asked to (1) create an account using their Semantic Scholar ID (users who do not have Semantic Scholar IDs can generate their interest models manually) in order to create their interest models, (2) interact with the application based on given tasks, and (3) take a closer look at the \textit{Why} and \textit{How} explanations provided by the system. Following a think-aloud approach, participants were also asked to say anything that comes to their mind during each interaction. After that, we conducted semi-structured interviews to gather in-depth feedback. The interviews took place online and were recorded with the consent of the participants. They lasted 10 to 15 minutes with the following questions: 
\textit{
(1) What do you like the most about the provided (Why / How) explanations?
(2) What do you like the least about the provided (Why / How) explanations?
(3) Which of the provided explanations (Why / How) helped you more to make a decision? Why?
(4) Which explanation (Why / How) is sufficient for you to make a decision?
(5) Which explanation (Why / How) do you prefer? Why?
(6) Which explanation (Why / How) gives you a better sense of transparency of the recommender system? Why?
(7) Which explanation (Why / How) gives you a better sense of trust in the recommender system? Why?
(8) Do you have any suggestions to improve the system?
}
After the semi-structured interviews, participants were also invited to fill out a questionnaire containing questions regarding usability aspects and attitudes towards the RS, based on the ResQue evaluation framework \cite{pu2011user}. To note that by using the ResQue framework, we are not aiming at conducting a quantitative evaluation and generalizing our conclusions, but rather to use participants' answers to the ResQue questionnaire as a starting point to collect their opinions towards the RS, which are then explored in-depth through our qualitative study.

\subsection{Analysis and results}
\begin{figure}[!ht]
\centering
\includegraphics[height=0.5 \textwidth]{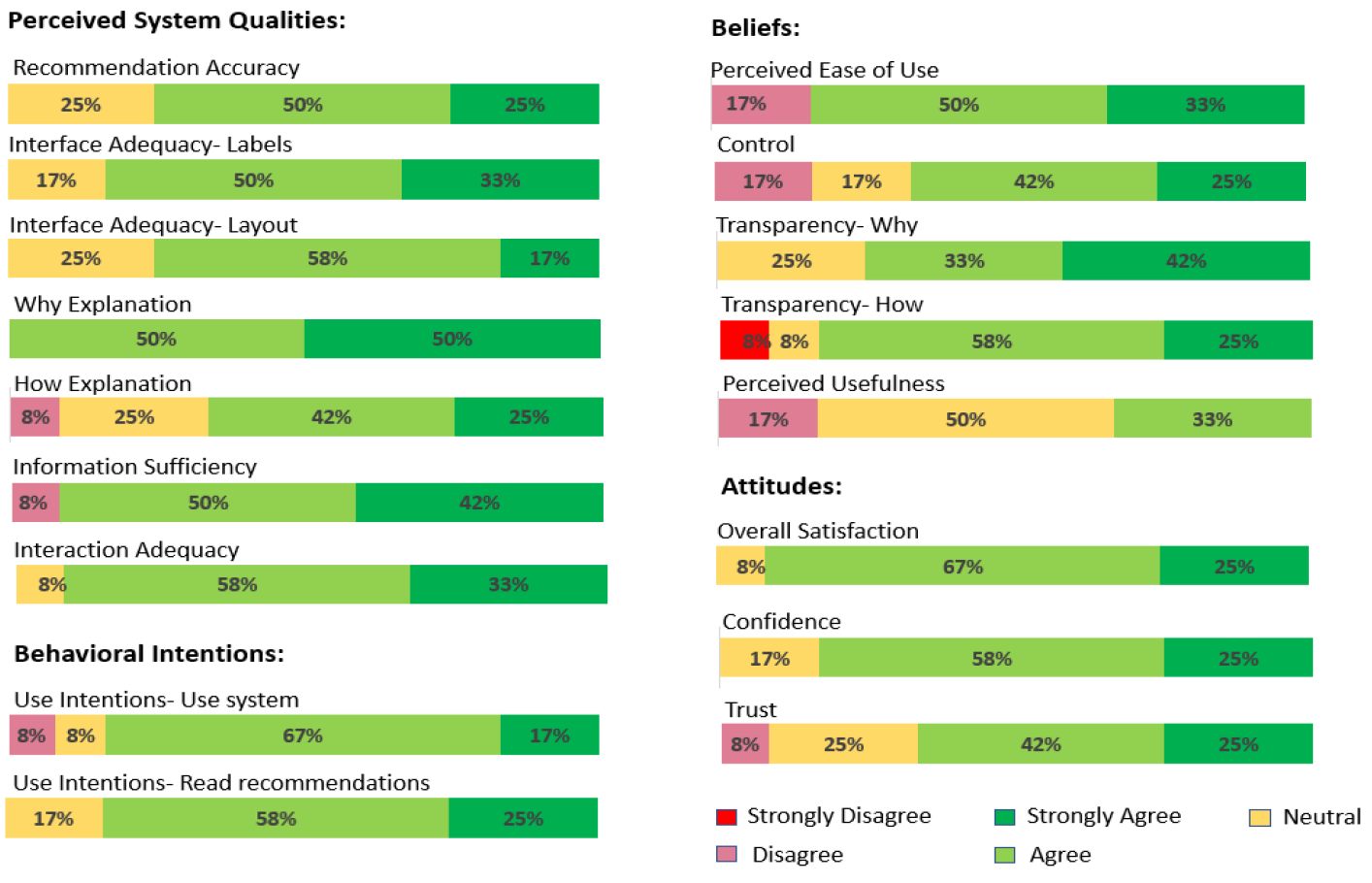}
\caption{Results from the ResQue questionnaire}
\label{fig:ResQue}
\end{figure}

The results of the ResQue questionnaires are summarized in Figure \ref{fig:ResQue}. We conducted a qualitative analysis of the moderated think-aloud sessions and the semi-structured interviews to gain further insights into the reasons behind the individual differences in the perception of the RS in terms of the \textit{Why} and \textit{How} explanations. We followed the instruction proposed by \citet{braun2006using} to code the data and identify patterns to organize the codes into meaningful groups. Notes and transcripts of the interview recordings were made for the analysis. The analysis was rather deductive as we aimed to find additional explanations for the users' opinions towards the three themes/goals that we are addressing with our research question, namely \textit{Transparency}, \textit{Trust}, and \textit{Satisfaction}.
\subsubsection{Transparency} 
\begin{figure}[!ht]
\centering
\includegraphics[height=0.3 \textwidth]{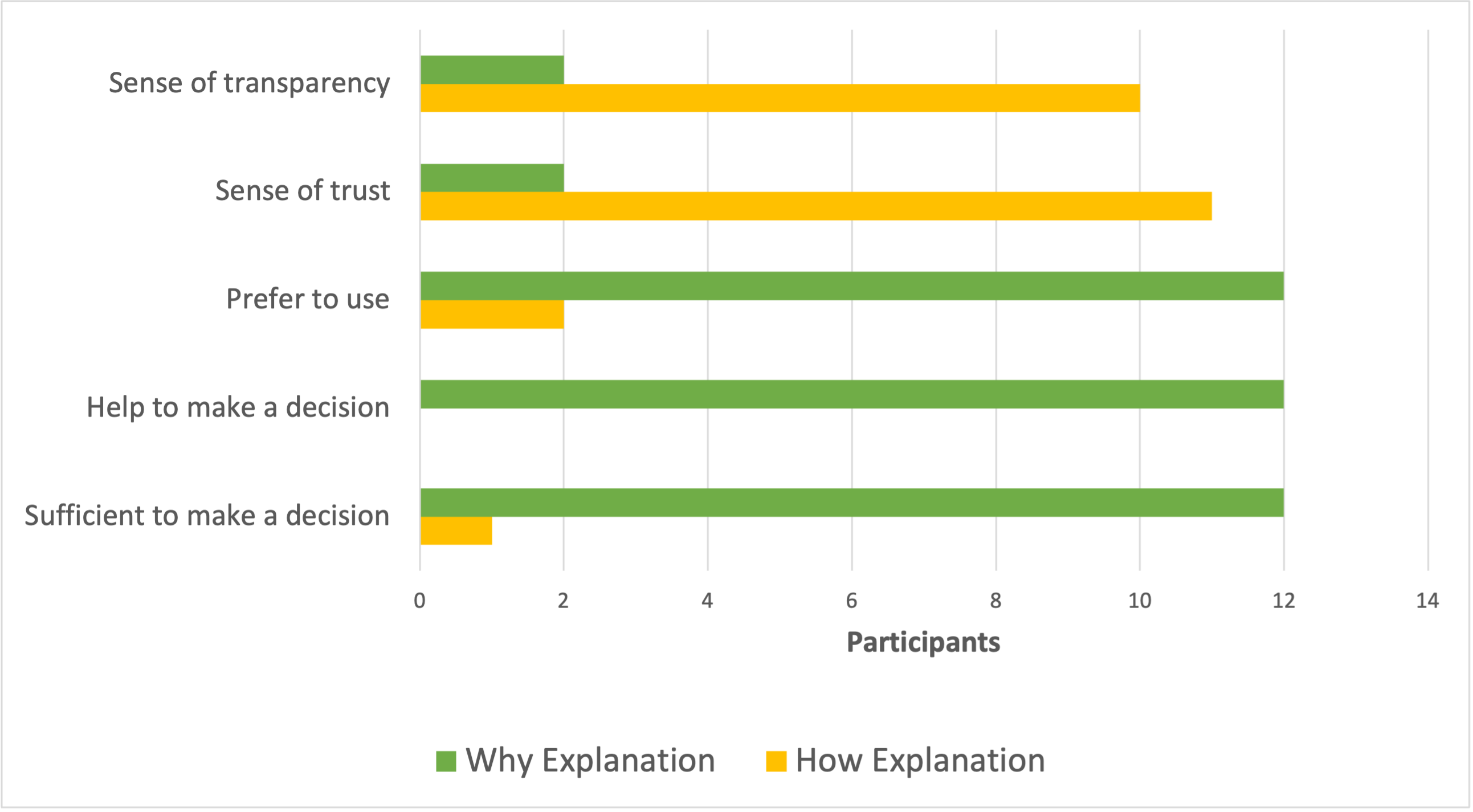}
\caption{Overall user experience with the \textit{Why} and \textit{How} explanations}
\label{fig:overall}
\end{figure}

This theme concerns the perception of the \textit{Why} and \textit{How} explanations in terms of transparency. In this regard, ten participants (respectively nine participants) stated that the \textit{How} explanation (respectively the \textit{Why} explanation) had an overall good effect on the transparency of the system (see Figure \ref{fig:ResQue}). When we concretely asked about which explanation gave them a better sense of transparency of the RS, the majority of participants agreed that they perceived the RS as more transparent through the \textit{How} explanation, as the system’s inner working was evident to them (see Figure \ref{fig:overall}). For instance, participant \textbf{P6} mentioned that \textit{"How explanation shows me the process of the system and lets me know what is happening behind it. Moreover, I can choose how much information I want to see”}. Also, \textbf{P7} reported \textit{“As an engineer, I always use mathematical formulas. So, if I know how the similarity scores are calculated, the system is more transparent for me”}. On the other hand, only two participants reported that the system was transparent because of the \textit{Why} explanation, as it provided enough information for them to understand the RS functionality. For instance, \textbf{P2} pointed out that \textit{“Why explanation increases the system’s transparency to me. It highlights the extracted keywords from the paper’s abstract with the same color as the relevant interest and also displays a similarity score to each one of them”}. These two participants further claimed that, by contrast, the \textit{How} explanation is difficult to understand as it contains very technical details that are difficult for non-computer scientists to comprehend.
\subsubsection{Trust} 
Regarding the perceived trust in the RS, eight participants found that the \textit{Why} and \textit{How} explanations had a positive impact on trust in general (see Figure \ref{fig:ResQue}). When we specifically asked which explanation gave them a better sense of trust in the RS, most participants agreed that the \textit{How} explanation increased their trust in the system, because the system's inner working was transparent to them (see Figure \ref{fig:overall}). For instance, \textbf{P6} indicated that \textit{"I would trust the system because it provides me exactly what is happening there via the how explanation"}. On the other hand, two participants expressed that the \textit{Why} explanation made them feel more confident in the RS and that the {How} explanation was overwhelming for them, which hurts their understanding and trust in the system. For instance, \textit{P4} mentioned \textit{"What does a chemical engineer have to do with this flow chart? I mean, I have some programming background and it can be interesting for me, but it is complicated for people who do not know how the algorithm works"}. 
\subsubsection{Satisfaction}
Most participants expressed high overall satisfaction with the RS (see Figure \ref{fig:ResQue}). As pointed out by \citet{tintarev2007survey}, satisfaction can also be measured indirectly, measuring user loyalty. Thus, users’ use intentions can be seen as an indirect measure of loyalty and satisfaction with the system. In this regard, the majority of participants expressed their intention to use the system in the future or read the publications recommended by the system (see Figure \ref{fig:ResQue}). Moreover, beside the satisfaction with the system as a whole, one can also measure the user’s quality perception of the explanations themselves as an indication of user's satisfaction with the system \cite{tintarev2012evaluating,gedikli2014}.    
Figure \ref{fig:ResQue} shows that the system’s ratings related to the perceived system qualities, including recommendation accuracy, interface adequacy, explanation quality, information sufficiency, and interaction adequacy are relatively high, indicating that the overall satisfaction and user experience are positive toward having the \textit{Why} and \textit{How} explanations in the RS (see Figure \ref{fig:ResQue}). All participants showed high satisfaction with the \textit{Why} explanation. For instance, \textbf{P4} expressed \textit{“I like the simplicity and how things explain themselves in the why explanation. No tutorials are needed. Even if I didn't really watch the video that you provided earlier, I still would be able to get to the same conclusion"}. For the \textit{How} explanation, eight users mentioned that they were satisfied with it because they could see the amount of information that they preferred to see. In addition, they appreciated the feature of providing tooltips displayed on each node of the flow chart to provide additional information. Participants liked that they could learn how the system works: \textit{“\textbf{P3}: I am able to follow the system process in the How explanation from both the interest model and the publication side”}, as well as that the How explanation is personalized to their individual data: \textit{“\textbf{P1}: It’s great to see my own data used to explain how the system works”}. On the other hand, four participants did not speak very confidently about their satisfaction with the \textit{How} explanation. These participants reported complexity as the main reason: \textit{“\textbf{P6}: "too technical”} and \textit{“\textbf{P12}: "might be overwhelming”}. \citet{nunes2017systematic} noted that satisfaction is not considered as a single goal, but can be split into sub-goals of ease to use and usefulness. Figure \ref{fig:ResQue} shows that perceived ease of use of the \textit{Why} and \textit{How} explanations scored high. On the other hand, perceived usefulness received relatively lower ratings. All participants reported that they found the \textit{Why} explanation useful, but they had diverse opinions regarding the usefulness of the \textit{How} explanation. In the interview session, we gathered feedback concerning the usefulness of the \textit{Why} and \textit{How} explanations, the situations where each explanation could be used, and the usage frequency for each explanation. As shown in Figure \ref{fig:overall}), the \textit{Why} explanation was perceived as the more effective explanation by the majority of participants. They assumed that this explanation was helpful and sufficient for them to make a decision on whether the recommended publication was relevant to them or not. Consequently, they would prefer to use the \textit{Why} explanation more frequently than the \textit{How} explanation. Most participants agreed that the \textit{How} explanation is an interesting option that they would use in some concrete situations, but not frequently. For instance, \textbf{P6} said \textit{"At a higher level, I want to know how the system works. I will click on the HOW button, but I would say not so frequently"}. Two participants (\textbf{P2} and \textbf{P7}) stated that they would look at the \textit{How} explanation only if they had difficulty understanding the \textit{Why} explanation. 
\section{Discussion}
The primary research question we address in this work is: What is the potential impact of \textit{Why} and \textit{How} visual explanations on the user perceptions regarding transparency, trust, and user satisfaction, when these two explanations are provided together in an explainable RS?

Most participants agreed that providing \textit{Why} and \textit{How} explanations in the RS had a positive impact on their perceived transparency of the system, which confirms earlier findings that incorporating explanation is essential to make RS more transparent \cite{herlocker2000explaining,tintarev2015explaining,nunes2017systematic,zhang2020explainable}. Moreover, our results showed that the \textit{How} explanation offered users a better sense of transparency of the RS, since it reveals the inner working of the system. This indicates that the \textit{How} explanation is the right choice, if the goal is to increase objective transparency. Our results further showed that for some users, the \textit{Why} explanation provided a better sense of transparency, as it provided enough information for them to understand the RS functionality, compared to the \textit{How} explanation which was difficult to understand. This suggests that the \textit{How} explanation increases the system’s objective transparency but is also associated with a risk of reducing the user-perceived transparency, depending on the user’s background knowledge. This confirms findings in previous studies showing that, for some users, it is enough to provide a \textit{Why} explanation to justify a recommendation output instead of revealing the inner working of the RS (e.g., \cite{gedikli2014,zhao2019users}. This further suggests that for assessing transparency in RS, it is necessary to view transparency as a multi-faceted concept and to differentiate objective transparency from user-perceived transparency \cite{gedikli2014,hellmann2022development}.

Regarding the perceived trust in the RS, most participants found that the system is trustworthy through the \textit{How} explanation, which is in line with findings in e.g., \cite{tintarev2012evaluating,kunkel2019let,pu2012evaluating,vig2009tagsplanations}, considering transparency as an important factor that contributes to users building trust in the RS, as it can enhance users' perceived understanding of the system. On the other hand, few participants identified the \textit{Why} explanation as more trustworthy, mainly because the \textit{How} explanation was overwhelming for them. Our findings imply a relationship between the user type (e.g., background knowledge) and the needed amount of information in an explanation. This is in line with the findings in e.g., \cite{guesmi2022explaining,chatti2022more,kouki2019personalized,millecamp2019,szymanski2021visual,martijn2022knowing}, showing that personal characteristics have an effect on the perception of RS explanations. Our findings also confirm the results of previous research on explainable recommendation and XAI showing that the detailed explanation does not automatically result in higher trust because the provision of additional explanations increases cognitive effort \cite{kulesza2015principles,kulesza2013too,yang2020visual,zhao2019users,kizilcec2016much,chatti2022more}. This line of research stresses that there is a trade-off between the amount of information in an explanation and the level of perceived trust users develop when interacting with the system and concludes that designing for trust requires balanced system transparency: “not too little and
not too much" \cite{kizilcec2016much} and "be sound", “be complete” but “don’t overwhelm” \cite{kulesza2015principles,kulesza2013too}. In summary, it is vital to find an optimal level of transparency that will generate the highest level of users’ trust in RS \cite{zhao2019users} and to provide 
personalized explanations with the right level of detail by tailoring the explanation intelligibility type to the user's context, i.e., goals and personal characteristics \cite{chatti2022more,ain2022multi}.  

Overall, our results show that providing \textit{Why} and \textit{How} explanations together within an RS leads to increased transparency, trust, and overall satisfaction. This is in line with earlier studies which found that the user’s overall satisfaction with an RS is assumed to be strongly related to transparency and trust. \citet{gedikli2014}, for example, reported results from experiments with different explanations clearly showing that transparency – independent of the used explanation – has a significant positive effect on user satisfaction. Similarly, \citet{balog2020measuring} found that satisfaction is positively correlated with transparency and trust. Regarding the users' perceptions of the \textit{Why} and \textit{How}, we observed a trade-off between transparency and trust on the one hand and satisfaction on the other hand "transparency/trust vs. satisfaction". Concretely, \textit{How} explanations can lead to higher transparency and trust, but lower satisfaction. \textit{Why} explanations, by contrast, are perceived as less transparent and trustworthy, but can contribute to increased satisfaction. One possible implication of this finding is to provide \textit{Why} explanations (by default) and \textit{How} explanations (on-demand) in order to increase transparency, trust, and overall satisfaction with the RS, at the same time. Furthermore, the fact that, while the \textit{How} explanation offered users a better sense of transparency and trust, the \textit{Why} explanation had a higher positive impact on users' satisfaction with the RS, confirms that there are inter-dependencies between explanation goals and intelligibility types and that different intelligibility types can be used for different explanation goals \cite{lim2013evaluating,lim2009why,lim2019these,liao2020questioning}. In our study, we identified specific pathways mapping the use of \textit{Why} and \textit{How} explanations back to the explanation goals of transparency, trust, and satisfaction. Concretely, While \textit{How} explanations can be mapped back to transparency and trust, \textit{Why} explanations are more linked to satisfaction. 

In general, participants were more satisfied with the \textit{Why} explanation which was also perceived as relatively simple, yet more effective than the \textit{How} explanation to make a decision. These observations are in line with those made in previous work. \citet{herlocker2000explaining}, for example, found in their study that the most satisfying explanations were simple and conclusive methods, such as stating the neighbors’ ratings, and that complex explanations such as a full neighbor graph scored significantly lower. Similarly, \citet{putnam2019exploring} and \citet{conati2021toward} reported that students want to know why more than they want to know how AI-driven hints are provided in intelligent tutoring systems (ITS). This suggests that, if an explainable RS only provides a single explanation, the focus should rather be on providing a \textit{Why} explanation (i.e., justification). 
\section{Limitations}
As a first analysis of the impact of \textit{Why} and \textit{How} explanations on users' perceptions, when these two explanations are provided together in an explainable RS, this study is not without limitations. First, we performed this analysis in a single domain. It must be verified that our findings transfer to domains beyond scientific literature RS. From the perspective of evaluation, we conducted a qualitative user study with 12 participants. Therefore, the results of the study should be interpreted with caution and cannot be generalized. A quantitative user study with a larger sample would probably have yielded more significant and reliable results.

\section{Conclusion and Future Work}
In this paper, we identified relationships between the \textit{Why} and \textit{How} explanation intelligibility types and the explanation goals of \textit{justification} and \textit{transparency}. We followed the Human-Centered-Design (HCD) approach and leveraged the What-Why-How visualization framework to systematically design \textit{Why} and \textit{How} visual explanations and provide them side-by-side in the transparent Recommendation and Interest Modeling Application (RIMA). Further, we presented a qualitative investigation of users' perceptions of \textit{Why} and \textit{How} explanations in terms of transparency, trust, and satisfaction. As a high level summary, we found qualitative evidence confirming that \textit{Why} and \textit{How} explanations have different effects on users and that the choice of these explanation intelligibility types depends on the explanation goal and user type. Moreover, we identified potential dependencies and trade-offs between the different explanation goals of transparency, trust, and satisfaction, when \textit{Why} and \textit{How} explanations are provided together in an explainable recommender system (RS).  

This work contributes to the literature on user-centered explanations. While we are aware that our results are based on one particular RS and the results cannot be generalized, we are confident that they represent a necessary step towards a richer understanding of the relationships between explanation intelligibility types and explanation goals in explainable RS. Future directions concern the generalization of findings to other application domains. Moreover, we plan to validate our findings through quantitative research to investigate in more depth the effects of providing \textit{Why} and \textit{How} explanations together on the perception of and interaction with explainable RS, with different user groups and in different contexts. Furthermore, we consider identifying more pathways mapping the different explanation intelligibility types back to the different explanation goals to be an important next step in our future work.
\bibliographystyle{ACM-Reference-Format}
\bibliography{sample-base}

\end{document}